\documentclass[12pt]{iopart}
\usepackage{graphicx}

\begin{document}

\title{Image formation in weak gravitational lensing by tidal charged black
holes }
\author{Zsolt Horv\'{a}th$^{1,2\dag }$, L\'{a}szl\'{o} \'{A}rp\'{a}d Gergely$%
^{1,2\ddag }$, David Hobill$^{3\ast }$}

\address{
$^{1}$ Department of Theoretical Physics, University of Szeged, Tisza L krt 84-86, Szeged 6720, Hungary \\
$^{2}$ Department of Experimental Physics, University of Szeged, D\'{o}m t\'{e}r 9, Szeged 6720, Hungary\\ 
$^{3}$ Department of Physics and Astronomy, University of Calgary, Calgary Alberta T2N 1N4 Canada\ \\
{\small 
$^{\dag }$zshorvath@titan.physx.u-szeged.hu; 
$^{\ddag }$gergely@physx.u-szeged.hu;
$^{\ast}$hobill@phas.ucalgary.ca
}}

\begin{abstract}
We derive a generic weak lensing equation and apply it for the study of
images produced by tidal charged brane black holes. We discuss the
similarities and point out the differences with respect to the Schwarzschild
black hole weak lensing, to both first and second order accuracy, when
either the mass or the tidal charge dominates. In the case of mass dominated
weak lensing, we analyze the position of the images, the magnification
factors and the flux ratio, as compared to the Schwarzschild lensing. The
most striking modification appears in the flux ratio. When the tidal charge
represents the dominating lensing effect, the number and orientation of the
images with respect to the optical axis resembles the lensing properties of
a Schwarzschild geometry, where the sign associated with the mass is
opposite to that for the tidal charge. Finally it is found that the ratio of
the brightness of images as a function of image separation in the case of
tidal charged black holes obeys a power-law relation significantly different
from that for Schwarzschild black holes. This might provide a means for
determining the underlying spacetime structure.
\end{abstract}

\maketitle

\section{Introduction}

Gravitational lensing has become a useful tool in measuring certain
properties of gravitational fields ever since the beginnings of general
relativity. While the initial observations of gravitational lensing were
used to verify the theoretical predictions of general relativity, it has now
been employed to study the large scale structure of the Universe, to
determine behaviour of compact stellar objects and to search for dark matter
candidates. In what follows we propose the idea that gravitational lensing
might also be used to determine which among various gravitational theories
is correct. It is already well known that the predictions of bending angles
computed from Newtonian gravity compared to those obtained from Einstein's
gravity differ by a factor of two. Therefore using gravitational lensing a
means for exploring the differences between competing gravitational theories
may well provide a technique that can be employed to determine the
dimensionality of spacetime, or the coupling of matter and fields to the
gravitational field or even to distinguishing among different formulations
of gravity theories.

In general the path taken by photons in a gravitational field provide a
number of different effects that can be measured using current
state-of-the-art telescopes. The first and most obvious one is the
production of multiple images and the relative separation of those images.
Secondly gravitational lensing can produce a change in brightness of the
images depending on how much bending a group of initially parallel rays
undergoes. Finally if the source or lensing object has a time dependency,
the changes in arrival times of light signals can provide a very accurate
measurement of some spacetime properties.

In this paper, we discuss the formation of images (i.e. their location and
brightness) for some black holes that are predicted to exist in
5-dimensional brane-world theories. By computing the bending angles and
image brightness changes that can occur due to the passage of photons past
the objects described by black hole solutions of the theory, we should be
able to determine enough of the properties of the lensing object to
distinguish a general relativistic black hole from that predicted by an
alternative theory.

Brane-world models have standard-model matter confined to a 3+1 dimensional
hypersurface, and gravity acting in a higher-dimensional non-compact
space-time. Such models have attracted much attention in recent years, both
as candidate theories meant to solve the hierarchy problem and predicting
modified cosmological evolution \cite{BDEL}. Higher codimension branes were
considered in connection with conical singularities \cite{BGNS}, \cite%
{deRham}. For a codimension-one brane-world, matter is generated on the
brane by the junction conditions though the brane representing a
discontinuity in the extrinsic curvature \cite{MaartensLivRev}- \cite%
{VarBraneTensionPRD}. Although the early expectation to replace dark energy
was not met, codimension-one brane-worlds still can produce alternative
explanations for dark matter \cite{MakHarko}-\cite{Pal}. Therefore the study
of localized matter configurations, in particular black hole solutions
admitted in brane-world theories became important.

Analytic black hole solutions include six-dimensional locally Schwarzschild
solutions \cite{Kaloper}; static five-dimensional black holes localized on
the brane, with the horizon decaying in the extra dimension and generated by
energy condition violating shells \cite{KantiTamvakis}, \cite%
{KantiOlasagastiTamvakis}; and black holes with a radiating component in the
extra dimension \cite{EmparanFabbriKaloper}. Numerically small brane black
holes (compared to the five-dimensional curvature) were shown to exist as
five-dimensional Schwarzschild solutions \cite{KudohTanakaNakamura} in
asymptotically five-dimensional Anti de Sitter space-time.

The four-dimensional Schwarzschild metric can be also extended into the
fifth dimension as a black string \cite{ChRH}, which due to the
Gregory-Laflamme instability \cite{GL} can pinch off, leading to a black
cigar metric \cite{Gregory} (although under very mild assumptions, classical
event horizons will not pinch off \cite{HorowitzMaeda}). Gravity wave
perturbations of such a black-string brane-world were computed in Ref. \cite%
{SCMlet}.

The perturbative analysis of the gravitational field of a spherically
symmetric source in the weak field limit in the original Randall-Sundrum
setup (Schwarzschild black hole on a brane embedded in Anti de Sitter
five-dimensional space-time) has shown corrections to the Schwarzschild
potential scaling as $r^{-3}$ \cite{RS}, \cite{GarrigaTanaka} and \cite%
{Giddings}. However if the Schwarzschild black hole is embedded in another
higher-dimensional space-time, this scaling would not apply \cite%
{BlackString}. Both weak \cite{KarSinha}, \cite{MM} and strong \cite{Whisker}
gravitational lensing of various brane black holes were discussed, the topic
being reviewed in Ref. \cite{MajumdarMukherjee}.

The effective Einstein equation on the codimension-one brane admits a
spherically symmetric vacuum solution characterized by two parameters: the
mass $m$ and a tidal charge $q$, the latter arising from the Weyl curvature
of the 5-dimensional space-time in which the brane is embedded \cite{tidalRN}%
:%
\begin{eqnarray}
ds^{2} &=&-f\left( r\right) dt^{2}+f^{-1}\left( r\right) dr^{2}+r^{2}\left(
d\theta ^{2}+\sin ^{2}\theta d\varphi ^{2}\right) ~,  \nonumber \\
\ f &=&1-\frac{2m}{r}+\frac{q}{r^{2}}~.  \label{tidal}
\end{eqnarray}%
For $q\leq m^{2}$ this represents a black hole with horizons given by $%
r_{\pm }=m\pm \left( m^{2}-q\right) ^{1/2}$. For $q<0$ only $r_{+}\,$\ is
positive, therefore there is only one horizon. For $q=0$ the line element
describes the Schwarzschild metric and for $0<q\leq m^{2}$ it is formally
identical to the general relativistic Reissner-Nordstr\"{o}m electro-vacuum
solution with electric charge $Q=q^{1/2}$. In the limit $q=m^{2}$ the metric
becomes extremal, such that the two horizons coincide. For $q>m^{2}$ there
is no horizon at all. The metric is singular at $r=0$, thus it describes a
naked singularity. It should be noted that a negative tidal charge
strengthens gravity (the horizon is outside the Schwarzschild radius $2m$),
such that $q<0$ contributes to the localization of gravity on the brane. A
positive tidal charge weakens gravity, both horizons lying below the
Schwarzschild radius for $0<q\leq m^{2}$, and obstructing the apparition of
a horizon at all for $q>m^{2}$.

Observations on light deflection could in principle constrain both the
lensing black hole parameters and the underlying gravitational theory. The
lensing properties of a Schwarzschild geometry were thoroughly investigated
in Refs. \cite{VE}, \cite{Virbhadra}. Recently in Ref. \cite{tidalDeflection}
the deflection angle of light rays passing near the tidal-charged brane
black hole was computed up to the second order in the perturbation theory.
The Hamiltonian method gave identical results to the previously employed
Lagrangian method \cite{PADEU}. The light deflection was derived in terms of
the small parameters $\varepsilon :=m/b$ and $\eta :=q/b^{2}$. Here $b$\ is
the impact parameter, defined as the distance of the lensing object to the
straight line trajectory, which would occur in the absence of the lensing
object. The deflection angle to second order accuracy is 
\begin{equation}
\delta =4\varepsilon -\frac{3}{4}\pi \eta +\frac{15}{4}\pi \varepsilon
^{2}-16\varepsilon \eta +\frac{105}{64}\pi \eta ^{2}~.  \label{delta}
\end{equation}%
A first confrontation with Solar System measurements in Refs. \cite%
{tidalDeflection}, \cite{BoehmerHarkoLobo} led to constraints on $q$ and on
the brane tension $\lambda $.

Further exploring the consequences of the result (\ref{delta}), in the
present paper we study the formation of images and their magnification
factors in the tidal charged black hole geometry, focusing on the
similarities and differences with the purely general relativistic case.

In the case of weak lensing where the latter black hole solutions differ
from standard general relativistic black holes, we are able to develop a
perturbative computation of the bending angles and image brightnesses that
can be expected when the black hole masses produce the dominant effect.
However the restriction to objects with weak tidal charge (compared to mass)
is not necessary. As long as the impact parameter or distance of closest
approach for the light ray remains large compared to a natural length scale
associated with the tidal charge, the weak field limit is maintained and
tidally charged black hole light bending effects can also be computed.

In Section 2 we study the weak lensing under quite generic circumstances. We
derive a lens equation for weak lensing, which will allow for a study of the
lensing and image formation to second order accuracy in the chosen small
parameters. This lens equation is more generic than, and reduces to, the
Virbhadra-Ellis equation \cite{Virbhadra}-\cite{Bozza} in a properly defined
approximation. We give the explicit form of both lens equations applied to
the tidal charged black hole in the Appendix, in order to see at which order
the differences arise.

In Section 3 we employ the deflection angle (\ref{delta}) in our lens
equation and obtain an approximate equation for weak lensing of the tidal
charged black hole in a form of a cubic polynomial. For later comparison we
also review here the lensing by Schwarzschild black holes, which arises in
our formalism to first order when the mass parameter dominates over the
tidal charge parameter.

In Section 4 we discuss the corrections in the location of the images and
magnification factors, by including the contributions quadratic in the mass
parameter and linear in the tidal charge parameter. This analysis leads to
results that are similar to the general relativistic Reissner-Nordstr\"{o}m
black hole lensing, some of them discussed in Ref. \cite{Sereno}. Despite $q$
being similar to the square of the electric charge of the Reissner-Nordstr%
\"{o}m black hole, the case of a negative tidal charge is without
counterpart in general relativity.

Section 5 contains the analysis of the case, where the tidal charge
dominates over the mass, by keeping only the first order tidal charge
contribution. We determine the image locations and the magnification
factors. In Section 6 we discuss the second order corrections to the above
case, by including contributions linear in the mass and quadratic in the
tidal charge.

We summarize our findings in the Concluding Remarks. Our choice of units is
given by $G=1=c$.

\section{The lens equation}

\begin{figure}[tbp]
\begin{center}
\includegraphics[width=8cm]{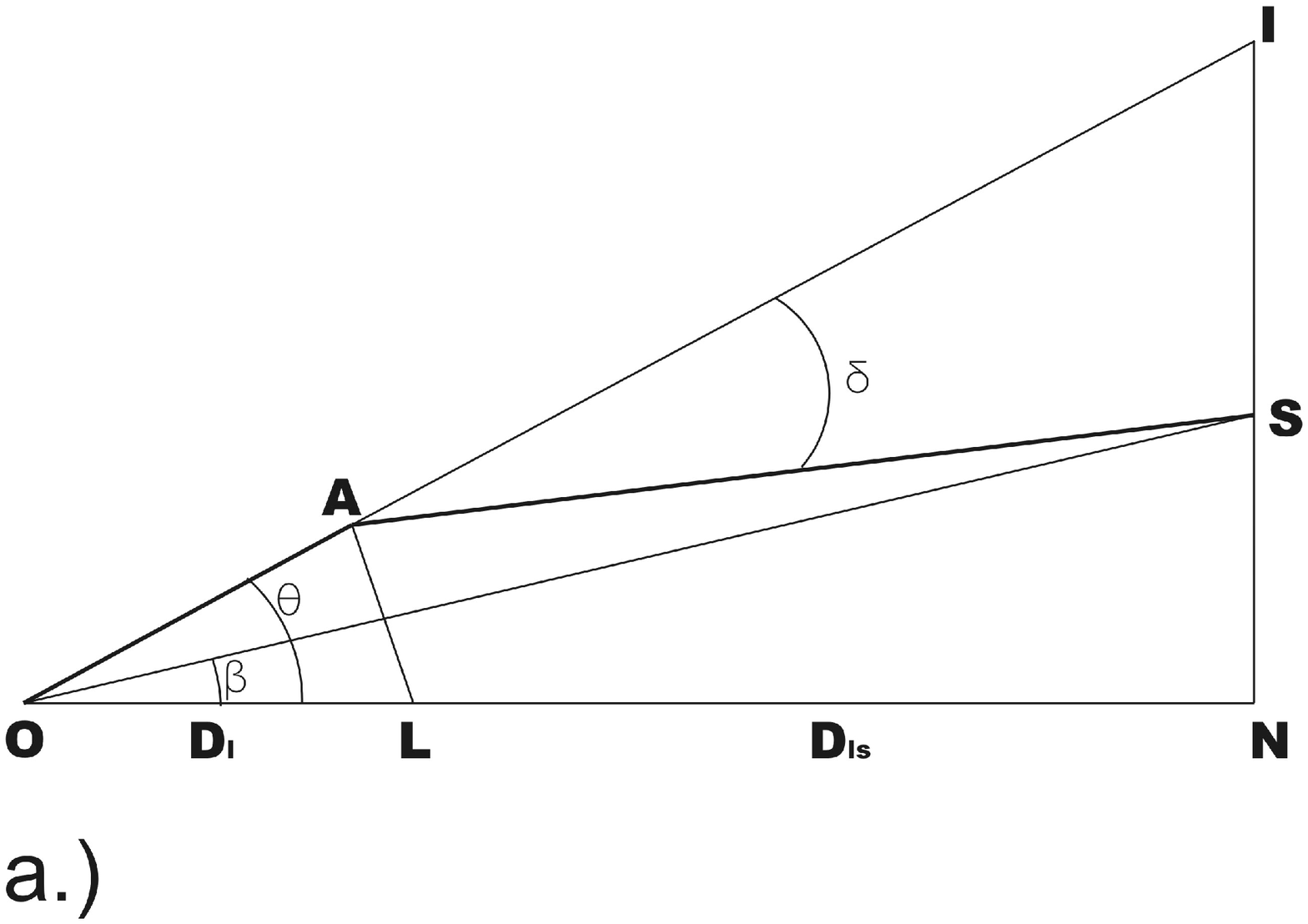} \vskip1cm %
\includegraphics[width=8cm]{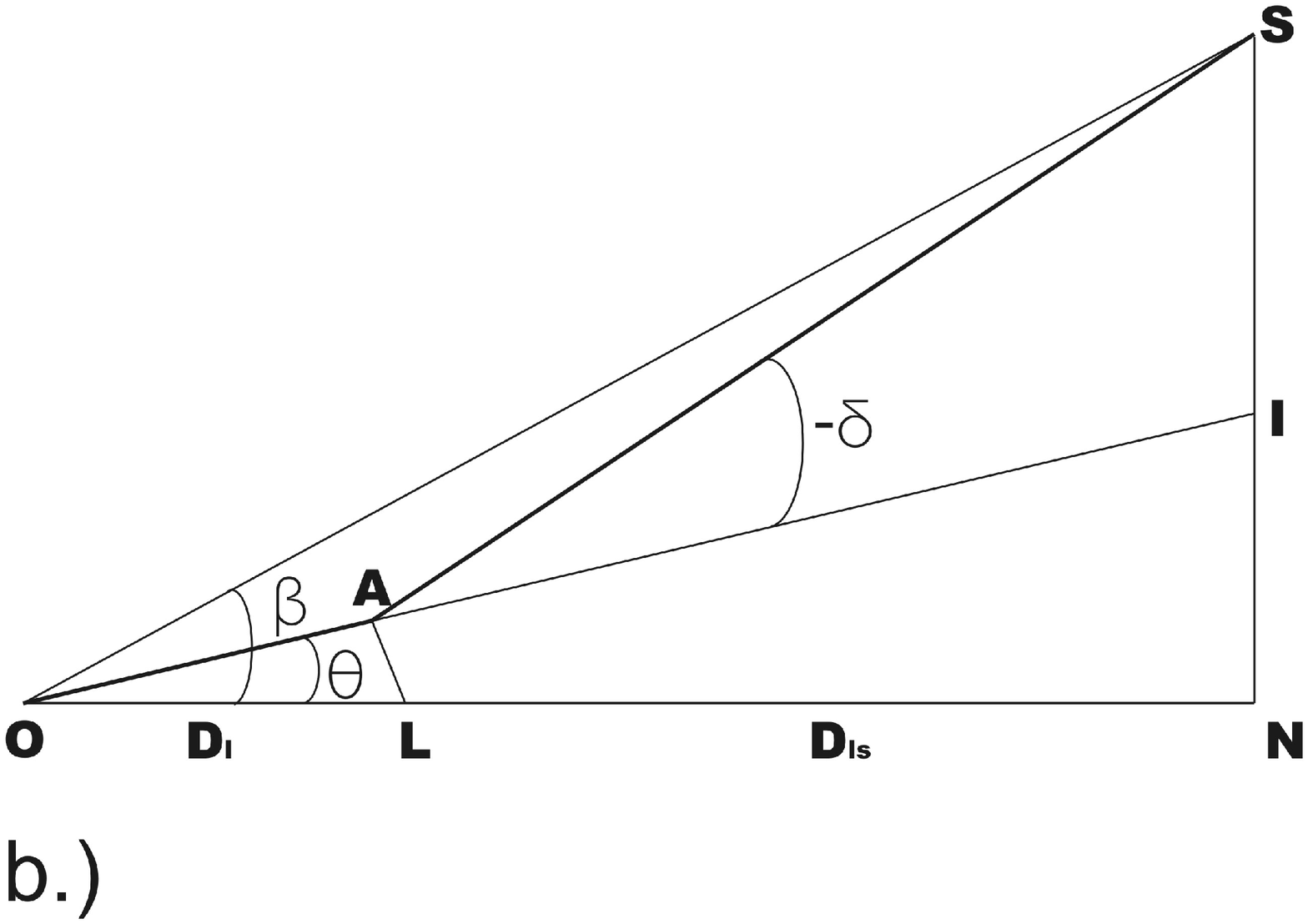} \vskip1cm %
\includegraphics[width=8cm]{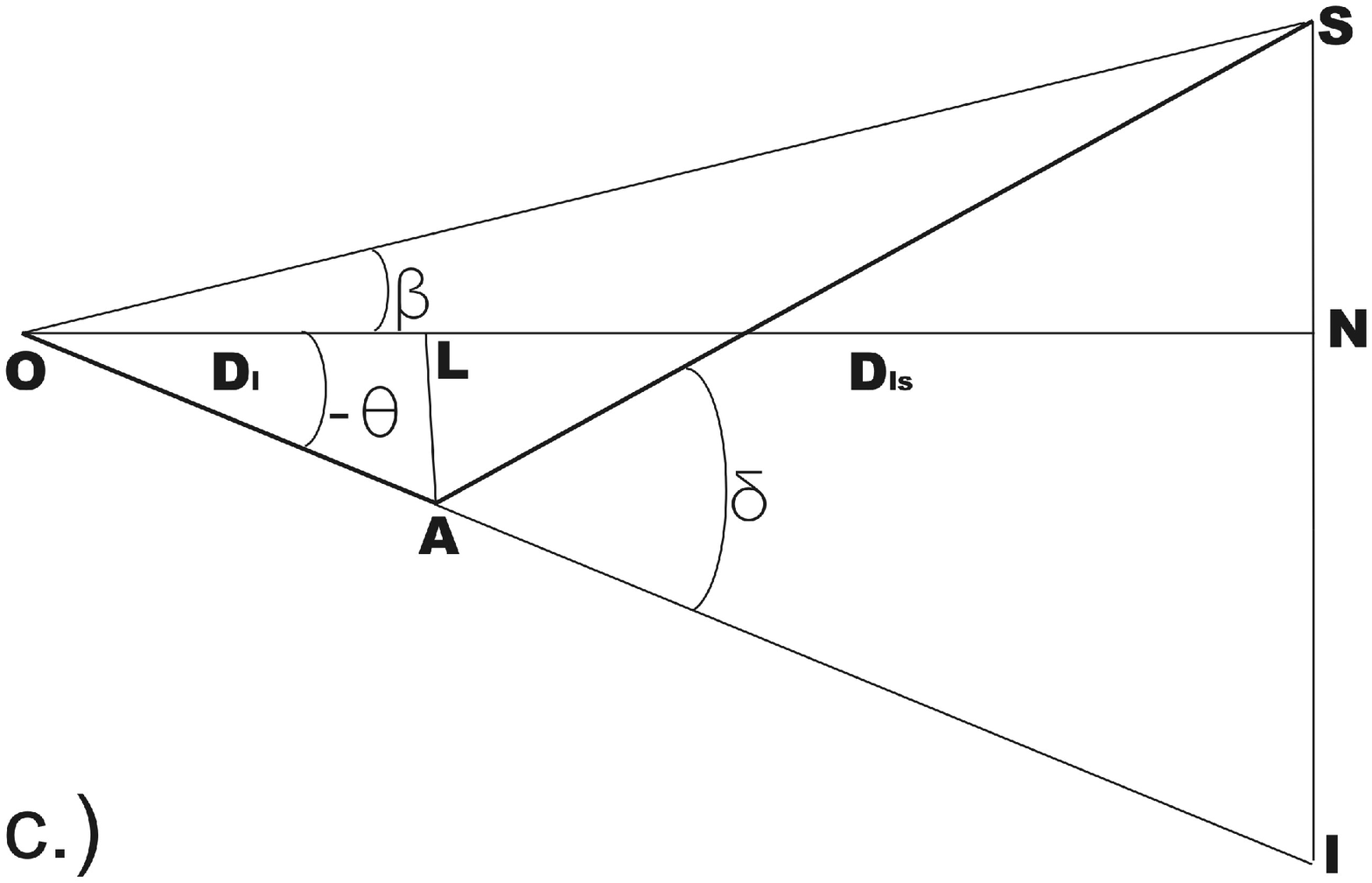}
\end{center}
\caption{Schematic representation of typical lensing configurations. The
light emitted from the source $S$, is deflected by the lensing object $L$
such that the observer at $O$ will see the image $I$. The angles $\protect%
\beta $ and $\protect\theta $ represent the angles spanned by the real and
apparent directions of the source with the line connecting the observer and
the lensing object. The light is bent towards the optical axis, while it
passes above (a) or below (c) the lensing object. Case (b) refers to a
repulsive interaction, which will also be encountered in this paper as a
limiting sub-case. On the figures we represent the (positive) length of the
arcs, expressed as $\protect\beta ,~\left\vert \protect\theta \right\vert =s%
\protect\theta $, $+\protect\delta $ for (a), (c) and $-\protect\delta $ for
(b).}
\label{fig1}
\end{figure}

In the lensing geometry the line connecting the lensing object ($L$) and the
observer ($O$) defines the optical axis. Relative to this axis the source
location ($S$) makes an angle $\beta =\widehat{LOS}$ from the optical axis,
chosen positive by convention (such that $S$ is always located
\textquotedblleft above" the optical axis, see Fig \ref{fig1}). Due to the
lensing effect however the source appears shifted away, and this is called
the image ($I$). The angle $\theta =\widehat{IOL}$ indicates the image
position and it can be either positive (for images above the optical axis)
or negative (for images below the optical axis). Let us denote $s=$sgn $%
\theta $, such that $\left\vert \theta \right\vert =s\theta $. Finally, the
deflection angle $\delta =\widehat{SAI}$ characterizes the change in the
direction of light due to the lensing object. We follow the convention that $%
\delta >0$ whenever the light is bent towards the optical axis and $\delta
<0 $ otherwise, cf. Ref \cite{Virbhadra}. Projecting the points $S$ and $I$
onto the optical axis ($\overline{OL}$) defines the distances $|\overline{LN}%
|$ or $D_{ls}$ from the lensing object and $|\overline{ON}|$ or $D_{s}$ from
the observer. The observer-lensing object distance is $D_{l}=D_{s}-D_{ls}$.

From Fig \ref{fig1} the angle $\widehat{ASI}=\pi /2+\left\vert \theta
\right\vert -\delta $. Then the sine theorem applied in the triangle $ASI$
reads%
\begin{equation}
\frac{\overline{NI}-s\overline{NS}}{\sin \delta }=\frac{\overline{OI}-%
\overline{OA}}{\cos \left( \delta -\left\vert \theta \right\vert \right) }~.
\label{sin1}
\end{equation}%
(We have taken into account the particularities of the configurations
indicated on Fig \ref{fig1} by including the sign $s$, whenever necessary.)
By multiplying the equation (\ref{sin1}) with $\sin \delta \cos \left(
\delta -\left\vert \theta \right\vert \right) /\overline{OI}$ we obtain%
\begin{equation}
\left( \frac{\overline{NI}}{\overline{OI}}-s\frac{\overline{NS}}{\overline{OI%
}}\right) \cos \left( \delta -\left\vert \theta \right\vert \right) =\left(
1-\frac{\overline{OA}}{\overline{OI}}\right) \sin \delta ~.  \label{l2}
\end{equation}%
We rewrite the left hand side by expressing $\sin \left\vert \theta
\right\vert ,$ $\cos \theta $ and $\tan \beta $ from the triangles $ONI$ and 
$ONS$. In the second term on the right hand side we rewrite $\overline{OI}%
=D_{s}\cos \theta $. We note that the point $A$ where the trajectory is bent
is defined such that $\widehat{OAL}=\widehat{SAL}=\left( \pi -\delta \right)
/2$. Then $\widehat{OLA}=\pi /2+\left\vert \theta \right\vert +\delta /2$,
and the sine theorem for the triangle $LAO$ gives%
\begin{equation}
\frac{\overline{OA}}{\cos \left( \frac{\delta }{2}-\left\vert \theta
\right\vert \right) }=\frac{D_{l}}{\cos \frac{\delta }{2}}~,  \label{sin2}
\end{equation}%
such that Eq. (\ref{l2}) becomes%
\begin{eqnarray}
0 &=&\frac{2D_{l}}{D_{s}}\cos \left( \frac{\delta }{2}-\left\vert \theta
\right\vert \right) \cos \left\vert \theta \right\vert \sin \frac{\delta }{2}
\nonumber \\
&&+\cos \left( \delta -\left\vert \theta \right\vert \right) \left( \sin
\left\vert \theta \right\vert -s\cos \left\vert \theta \right\vert \tan
\beta \right) -\sin \delta ~.  \label{LE}
\end{eqnarray}%
We would like to stress that this is an \textit{exact lens equation }in the
weak lensing approximation, as it was obtained exclusively by trigonometric
considerations, and no power series expansions of trigonometric functions
were applied.\footnote{%
A more general lens equation is also known for generic curved space-times 
\cite{FrittelliNewman}.} We will employ Eq. (\ref{LE}) to derive the
approximate lensing equation to the accuracy required by our approach.

Before doing so, we comment that in the particular configuration of $O,~L$
and $S$ being aligned (such that $\beta =0$) the above equation reduces to 
\begin{equation}
0=D_{l}\sin \left\vert \theta \right\vert +D_{ls}\sin (\left\vert \theta
\right\vert -\delta )~,  \label{ringeq}
\end{equation}%
the solution of which gives the angle for the formation of the Einstein ring:%
\begin{equation}
\left\vert \theta \right\vert =\arctan \frac{D_{ls}\sin \delta }{%
D_{l}+D_{ls}\cos \delta }~.
\end{equation}%
The usefulness of this expression is however limited, as $\delta $ is not an
observable quantity.

Secondly, we discuss the limit in which an approximate lens equation, one
that is frequently employed in the literature, arises. For this, we need to
assume $\overline{LA}\perp \overline{ON}$ (see Fig. 1 of Ref. \cite{VE}),
which introduces an error of order of the angle of deviation from
perpendicularity, which is $\theta -\delta /2$. Then to first order in $%
\epsilon :=\delta -2\left\vert \theta \right\vert $ the lens equation (\ref%
{LE}) becomes%
\begin{eqnarray}
0 &=&\tan \left\vert \theta \right\vert -s\tan \beta -\frac{D_{ls}}{D_{s}}%
\tan \left\vert \theta \right\vert -\frac{D_{ls}}{D_{s}}\tan \left( \delta
-\left\vert \theta \right\vert \right)  \nonumber \\
&&+\epsilon \tan \left\vert \theta \right\vert \left( s\tan \beta +\frac{%
D_{ls}}{D_{s}}\tan \left\vert \theta \right\vert \right) +\mathcal{O}\left(
\epsilon ^{2}\right) ~.
\end{eqnarray}%
The expansion in $\epsilon $ of the term $\tan \left( \delta -\left\vert
\theta \right\vert \right) $ contains the linear contribution $\epsilon
/\cos ^{2}\left\vert \theta \right\vert $. To linear order in $\epsilon $,
the last term can be approximated by replacing $\beta $ with its zeroth
order expression: $s\tan \beta =\tan \left\vert \theta \right\vert -2\left(
D_{ls}/D_{s}\right) \tan \left\vert \theta \right\vert $, obtaining $%
\epsilon \left( D_{l}/D_{s}\right) \tan ^{2}\left\vert \theta \right\vert $.
For small $\left\vert \theta \right\vert $, this last contribution can be
safely dropped as compared to $\epsilon /\cos ^{2}\left\vert \theta
\right\vert $. The terms to keep are%
\begin{equation}
0=\tan \left\vert \theta \right\vert -\tan \left( s\beta \right) -\frac{%
D_{ls}}{D_{s}}\left[ \tan \left\vert \theta \right\vert +\tan \left( \delta
-\left\vert \theta \right\vert \right) \right] ~.  \label{lens_Ellis}
\end{equation}%
With a change in notation where $\left\vert \theta \right\vert \rightarrow
\theta _{VE}$, $s\beta \rightarrow \beta _{VE}$ which corresponds to the
change in the convention of which angles are defined to be positive
(specifically $\beta \geq 0$ in our approach and $\theta _{VE}\geq 0$ in
Refs. \cite{Virbhadra}-\cite{Bozza}), the equation (\ref{lens_Ellis}) is
known as the Virbhadra-Ellis lens equation. From the way it arises in our
formalism, we can tell that this equation is valid for small angles $%
\left\vert \theta \right\vert $ and to first order accuracy in $\delta
-2\left\vert \theta \right\vert $. This quantity vanishes if S, L and O are
collinear ($\beta =0$) and L is midway between S and O. It is therefore
expected, that the predictions of the more exact lens equation (\ref{LE})
will differ from the predictions of Eq. (\ref{lens_Ellis}) in asymmetric
setups of S and O with respect to L.

Since we are interested in possibly higher order contributions,
characterizing the lensing by tidal charged brane black holes and naked
singularities with tidal charge, we need to improve the level of
approximation employed in Eq. (\ref{lens_Ellis}). Therefore we start our
investigations on the brane-world lensing process from the lens equation
derived here, Eq. (\ref{LE}) and we will employ a higher order expansion
than the one leading to Eq. (\ref{lens_Ellis}). We have also adopted the
convention $\beta >0$ as this will turn useful in the discussion of the
second-order effects.

The lens equation (\ref{LE}) presented here is obtained solely from the
geometry of the paths taken by both the deflected and undeflected light
rays. It relates the angles $\theta ,~\beta $ and $\delta $ to each other
through the relations among the trigonometric functions of those angles. The
lens equation in this form does not care whether the lensing carried out by
an optical instrument or a gravitational lens.

When the bending is due to gravitational effects, the deflection angle $%
\delta $ (characterizing the strength of the bending) can be derived from
the null geodesic equations. For weak lensing, the deflection angle can be
computed using a set of expansion parameters that characterize the geometry
of the black hole spacetime. In the cases studied in this paper we use the
mass and tidal charge divided by a characteristic length scale. The lens
equation can then be applied along with an expression for the bending angle $%
\delta$ to obtain the image locations $\theta$ given a source position $\beta
$.

The issue that arises is to ensure that the accuracy of the lens equation
matches the level of accuracy of the approximation in the deflection angle.
If the lens equation is accurate only to linear order in the angles and the
deflection angle is good to third order in the expansion parameters, then
there is a loss of accuracy in the former that no longer makes it suitable
for higher order computations. If this is not properly taken into account,
errors will arise. On the other hand two different expressions for the lens
equations can lead to indistinguishable results if they agree to within the
same order of approximation.

\section{Second order lens equation in the mass and tidal charge\label%
{2o_lens}}

In order to find the position of the images, first the expansion (\ref{delta}%
) of the deflection angle $\delta $ should be inserted in the lens equation (%
\ref{LE}), secondly the approximations following from the weak lensing
approach are carried out. An inconvenience in proceeding this way is that
the impact parameter $b=D_{l}\sin \left\vert \theta \right\vert $ entering
in the definitions of the small parameters $\varepsilon $ and $\eta $
depends on $\theta $. Therefore we introduce the alternative set of small
parameters 
\begin{equation}
\bar{\varepsilon}=\frac{m}{L}~,\quad \bar{\eta}=\frac{q}{L^{2}}~,
\end{equation}%
with $L=D_{s}D_{l}/D_{ls}$. A series expansion of Eq. (\ref{LE}) accurate to
second order in both small parameters gives%
\begin{equation}
0=L_{0}+\bar{\varepsilon}L_{10}+\bar{\eta}L_{01}+\bar{\varepsilon}^{2}L_{20}+%
\bar{\varepsilon}\bar{\eta}L_{11}+\bar{\eta}^{2}L_{02}~.  \label{LE1}
\end{equation}%
Here%
\begin{eqnarray}
L_{0} &=&\cos ^{2}\theta ~\left( \tan \left\vert \theta \right\vert -s\tan
\beta \right) ~,  \nonumber \\
L_{10} &=&-4\cos \left\vert \theta \right\vert \left( s\frac{L}{D_{l}}\tan
\beta +\cot \left\vert \theta \right\vert \right) ~,  \nonumber \\
L_{01} &=&\frac{3\pi }{4}\frac{L}{D_{l}}\cot \left\vert \theta \right\vert
\left( s\frac{L}{D_{l}}\tan \beta +\cot \left\vert \theta \right\vert
\right) ~,  \nonumber \\
L_{20} &=&\frac{L}{4D_{l}}\cot \left\vert \theta \right\vert \left[ s\frac{L%
}{D_{l}}\tan \beta \left( 32\cot \left\vert \theta \right\vert -15\pi
\right) -32-15\pi \cot \left\vert \theta \right\vert \right] ~,  \nonumber \\
L_{11} &=&-\frac{L^{2}}{D_{l}^{2}}\frac{\cos \left\vert \theta \right\vert }{%
\sin ^{2}\left\vert \theta \right\vert }\left[ s\frac{L}{D_{l}}\tan \beta
\left( 3\pi \cot \left\vert \theta \right\vert -16\right) -3\pi -16\cot
\left\vert \theta \right\vert \right] ~,  \nonumber \\
L_{02} &=&\frac{3\pi }{64}\frac{L^{3}}{D_{l}^{3}}\frac{\cos \left\vert
\theta \right\vert }{\sin ^{3}\left\vert \theta \right\vert }\left[ s\frac{L%
}{D_{l}}\tan \beta \left( 6\pi \cot \left\vert \theta \right\vert -35\right)
-6\pi -35\cot \left\vert \theta \right\vert \right] ~.
\end{eqnarray}%
The zeroth order contribution $L_{0}$ shows that without the black hole ($%
m=0=q$) there is no deflection, $\theta =\beta $.

\begin{figure}[tbp]
\begin{center}
\includegraphics[width=8cm]{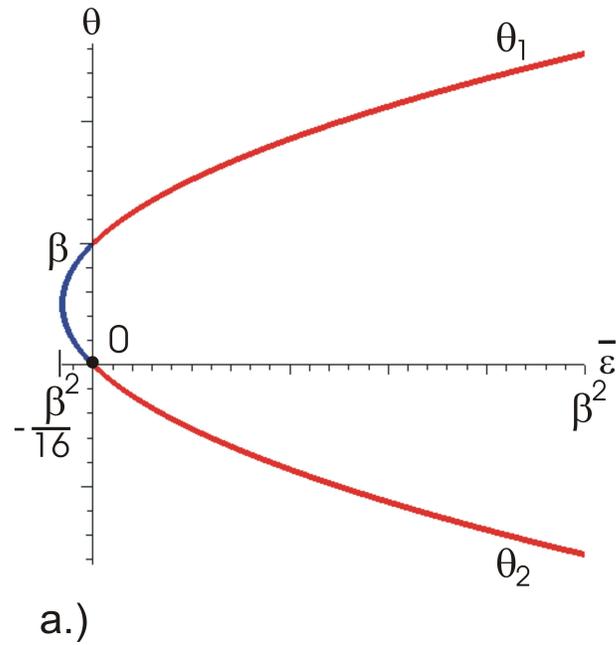} \vskip1cm %
\includegraphics[width=9cm]{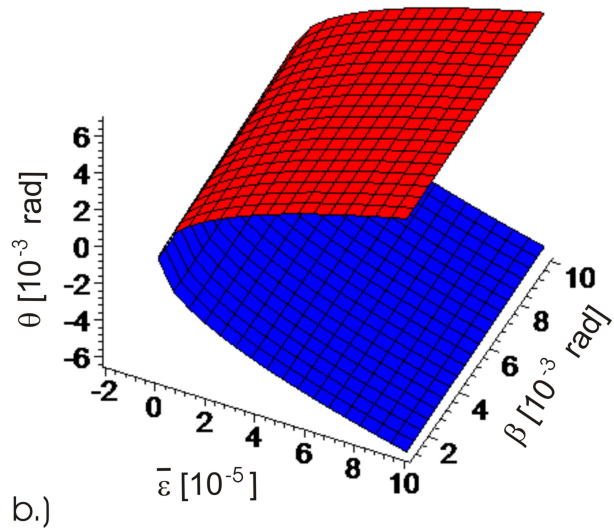}
\end{center}
\caption{ When the mass dominates over the tidal charge in the lensing, the
images arise as in the Schwarzschild case. (a) The position $\protect\theta $
(in units $\protect\beta $) of the images for Schwarzschild black holes is
characterized by the parameter $\bar{\protect\varepsilon}$ (in units $%
\protect\beta ^{2}$), $\protect\beta >0$ indicating the real position of the
source. For positive mass ($\bar{\protect\varepsilon}>0$) there are two
images, situated above and below the optical axis. For negative mass ($\bar{%
\protect\varepsilon}<0$) both images lie above the optical axis. The images
coincide for $\bar{\protect\varepsilon}=-\protect\beta ^{2}/16$. For each $%
\protect\beta $, the negative masses with $\bar{\protect\varepsilon}<-%
\protect\beta ^{2}/16$ do not allow for any image. (b) $\protect\theta $ as
function of $\bar{\protect\varepsilon}$ and $\protect\beta $. With
decreasing $\protect\beta $, the images shrink accordingly. At $\protect%
\beta =0,~$the angle $\protect\theta $ represents the angular radius of the
Einstein ring (therefore the $\protect\beta =0$ section of the surface is
symmetric with respect to $\protect\theta =0$).}
\label{fig2}
\end{figure}

Without the tidal charge and to first order in $\bar{\varepsilon}$ we obtain
the standard Schwarzschild lensing as 
\begin{equation}
0=\cos \left\vert \theta \right\vert \left[ \cos \left\vert \theta
\right\vert ~\left( \tan \left\vert \theta \right\vert -s\tan \beta \right)
-4\bar{\varepsilon}\left( \cot \left\vert \theta \right\vert +s\frac{L}{D_{l}%
}\tan \beta \right) \right] ~.
\end{equation}%
From the lensing geometry (see Fig \ref{fig1}, with the remark that the
involved distances are large and the deflection is relevant only for the
trajectories crossing nearby the lensing object) we can safely conclude,
that $\beta =\mathcal{O}\left( \theta \right) $. Assuming that $\bar{%
\varepsilon}=\mathcal{O}\left( \theta ^{2}\right) $ (we will see that in our
approach this condition is necessary for weak lensing), to leading order we
obtain%
\begin{equation}
0=~\theta ^{2}-\beta \theta -4\bar{\varepsilon}\equiv \mathcal{S}~,
\label{LE_Schw}
\end{equation}%
with the known solutions 
\begin{equation}
\theta _{1,2}=\frac{\beta \pm \sqrt{\beta ^{2}+16\bar{\varepsilon}}}{2}~.
\label{sg}
\end{equation}%
The position of the images is represented on Fig \ref{fig2}. With perfect
alignment of the source, lensing object and observer along the optical axis
we get the angular radius of the Einstein ring $\theta _{E}=2\bar{\varepsilon%
}^{1/2}$. This verifies the correctness of our assumption on the order of $%
\bar{\varepsilon}$.

For a hypothetical negative lensing mass there are still two images \cite%
{Cramer}, provided the discriminant stays non-negative, thus for $\beta
^{2}-16\left\vert \bar{\varepsilon}\right\vert \geq 0$. This time however
both\ images have a positive $\theta $. At equality the two images coincide.

For a lens with axial symmetry the magnification factor (the ratio of the
solid angle subtended by the image divided by the solid angle subtended by
the source) is given by \cite{Schneider}, \cite{BlandfordNarayan}%
\begin{equation}
\mu =\left\vert \frac{\theta }{\beta }\frac{d\theta }{d\beta }\right\vert ~,
\label{amplification}
\end{equation}%
where both $\theta $ and $\beta $ are small angles. For a Schwarzschild
lens, we substitute the images (\ref{sg}) and obtain 
\begin{equation}
\mu _{1,2}=\frac{1}{4}\left( \frac{1}{\beta }\sqrt{\beta ^{2}+4\theta
_{E}^{2}}+\frac{\beta }{\sqrt{\beta ^{2}+4\theta _{E}^{2}}}\pm 2\right) ~.
\label{amplificationS}
\end{equation}%
When $\beta \rightarrow 0$%
\begin{eqnarray}
\mu _{1,2} &=&\allowbreak \frac{\theta _{E}}{2\beta }\pm \frac{1}{2}+\frac{%
3\beta }{8\theta _{E}}+\mathcal{O}\left( \beta ^{2}\right) ~,  \nonumber \\
\frac{\mu _{1}}{\mu _{2}} &=&1+\frac{2\beta }{\theta _{E}}+\frac{2\beta ^{2}%
}{\theta _{E}^{2}}+\mathcal{O}\left( \beta ^{3}\right) ~,  \nonumber \\
\theta _{1,2} &=&\pm \allowbreak \theta _{E}+\frac{\beta }{2}\pm \frac{\beta
^{2}}{8\theta _{E}}+\mathcal{O}\left( \beta ^{3}\right) ~,  \label{smallbeta}
\end{eqnarray}%
thus the magnification factors diverge, while their ratio (the flux ratio) $%
\mu _{1}/\mu _{2}$ goes to unity. So far we have reproduced known results.

In what follows, we will discuss two novel applications:

A) The case when the tidal charge contributes to second order, thus $%
\mathcal{O}\left( \bar{\eta}\right) <\mathcal{O}\left( \bar{\varepsilon}%
\right) $. This case will be discussed to second order accuracy. (For
positive tidal charge, with the replacement $q\rightarrow Q^{2}$ we recover
Reissner-Nordstr\"{o}m lensing.)

B) The case when the tidal charge dominates, thus $\mathcal{O}\left( \bar{%
\varepsilon}\right) <\mathcal{O}\left( \bar{\eta}\right) $. Here for
simplicity first we go only to first order in $\bar{\eta}$, this being
formally equivalent to dropping all $\bar{\varepsilon}$ terms. Subsequently,
we will analyze the second order corrections.

In what follows, the black hole parameters will be related to $\mathcal{O}%
\left( \theta \right) $ by assumptions well justified case-by-case.
Equivalently, we will investigate the weak lensing properties of tidal
charged black holes in the corresponding ranges of its parameters.

\section{Mass dominated weak lensing to second order\label{2o_m}}

Under the assumptions A) of Section \ref{2o_lens} the terms $\bar{\varepsilon%
}\bar{\eta}$ and $\bar{\eta}^{2}$ can be dropped from the lens equation (\ref%
{LE1}). By taking as before $\bar{\varepsilon}=\mathcal{O}\left( \theta
^{2}\right) $, then $\bar{\eta}\leq \mathcal{O}\left( \bar{\varepsilon}%
^{2}\right) =\mathcal{O}\left( \theta ^{4}\right) $ and keeping only the
first and second order terms we obtain%
\begin{equation}
0=\frac{\mathcal{S}}{\theta }+s\gamma \frac{\bar{\eta}-5\bar{\varepsilon}^{2}%
}{\theta ^{2}}~,
\end{equation}%
or%
\begin{equation}
0=\theta ^{3}-\beta \theta ^{2}-4\bar{\varepsilon}\theta +s\gamma \left( 
\bar{\eta}-5\bar{\varepsilon}^{2}\right) ~.  \label{LE_m}
\end{equation}%
where we have introduced the notation%
\begin{equation}
\gamma =\frac{3\pi }{4}\frac{L}{D_{l}}>2.35~.  \label{gamma}
\end{equation}%
With $\bar{\eta}=\mathcal{O}\left( \theta ^{\geq 4}\right) $, the last term
of Eq. (\ref{LE_m}) represents a perturbation of the Schwarzschild lensing.
We look for solutions therefore in the form 
\[
\widetilde{\theta }=\theta _{1,2}\left[ 1+\mathcal{T}\left( \beta ,\gamma ,s,%
\bar{\varepsilon},\bar{\eta}\right) \right] ~, 
\]%
with $\mathcal{T}\theta _{1,2}$ a correction to the Schwarzschild images
located at $\theta _{1,2}$. The solutions are

\begin{equation}
\mathcal{T}_{1,2}=\frac{s\gamma \left( \bar{\eta}-5\bar{\varepsilon}%
^{2}\right) }{\theta _{1,2}\left( -3\theta _{1,2}^{2}+2\beta \theta _{1,2}+4%
\bar{\varepsilon}\right) }~.  \label{pert1}
\end{equation}%
We have employed that at $\theta _{1,2}$ the sign $s=\pm 1$. There are two
images, located at%
\begin{eqnarray}
\widetilde{\theta }_{1,2} &=&\theta _{1,2}\mp \mathcal{A}_{1,2}
\label{thetaScorr} \\
\mathcal{A}_{1,2} &\equiv &-s\theta _{1,2}\mathcal{T}_{1,2}=\frac{\gamma
\left( \bar{\eta}-5\bar{\varepsilon}^{2}\right) }{\beta \theta _{1,2}+8\bar{%
\varepsilon}}~.  \label{A}
\end{eqnarray}%
The second form of the expressions $\mathcal{A}_{1,2}$ eliminates the
quadratic term in $\theta _{1,2}$ appearing in Eq.~(\ref{pert1}) by using
Eq.~(\ref{LE_Schw}). By employing in the first form of $\mathcal{A}_{1,2}$
the explicit expressions for $\theta _{1,2}$, and introducing the variables $%
x^{\pm }=\widetilde{\theta }_{1,2}/\theta _{E}$, $x_{0}^{\pm }=\theta
_{1,2}/\theta _{E}$, $y=\beta /\theta _{E}$, also the notation $%
d_{RN}=\gamma \left( 5\bar{\varepsilon}-\bar{\eta}/\bar{\varepsilon}\right)
/4\theta _{E}$ we recover\footnote{%
Without the index 1 on the left hand side of the respective equation, which
is a typo.} Eq. (21) of Ref. \cite{Sereno}, derived for Reissner-Nordstr\"{o}%
m black holes. Our result however also covers the negative tidal charge
case, which has no analogue in general relativity.

At perfect alignment $\beta =0$, the second order accuracy Einstein ring
appears, as%
\begin{equation}
\widetilde{\theta }_{E}=\theta _{E}-\frac{\gamma }{8}\left( \frac{\bar{\eta}%
}{\bar{\varepsilon}}-5\bar{\varepsilon}\right) ~.
\end{equation}%
The second order correction to the Schwarzschild images and Einstein ring
computed here are of relative order $\mathcal{O}\left( \theta \right) $,
compared to the respective first order expressions, as expected.

\subsection{Magnification factors}

In this subsection we compute the corrections to the Schwarzschild
magnification (\ref{amplificationS}). For this we employ the second
expression (\ref{thetaScorr}) in Eq. (\ref{amplification}) and obtain%
\begin{equation}
\widetilde{\mu }_{1,2}=\mu _{1,2}\left( 1\mp \frac{\mathcal{A}_{1,2}}{\theta
_{1,2}}\frac{8\bar{\varepsilon}\pm \beta \mathcal{A}_{1,2}}{8\bar{\varepsilon%
}+\beta \theta _{1,2}}\right) \pm \frac{\theta _{1,2}}{\beta }\frac{\mathcal{%
A}_{1,2}\left( \theta _{1,2}\mp \mathcal{A}_{1,2}\right) }{\beta \theta
_{1,2}+8\bar{\varepsilon}}~.
\end{equation}

For $\mathcal{O}\left( \beta \right) \approx \mathcal{O}\left( \theta
\right) $ to leading order in $\theta $ and employing $\bar{\varepsilon}%
=\theta _{E}^{2}/4$ we are left with%
\begin{equation}
\widetilde{\mu }_{1,2}=\mu _{1,2}\left( 1\mp \frac{2\theta _{E}^{2}\mathcal{A%
}_{1,2}}{\left( 2\theta _{E}^{2}+\beta \theta _{1,2}\right) \theta _{1,2}}%
\right) \pm \frac{\mathcal{A}_{1,2}\theta _{1,2}^{2}}{\beta \left( 2\theta
_{E}^{2}+\beta \theta _{1,2}\right) }~.
\end{equation}

\begin{figure}[tbp]
\begin{center}
\includegraphics[width=10cm]{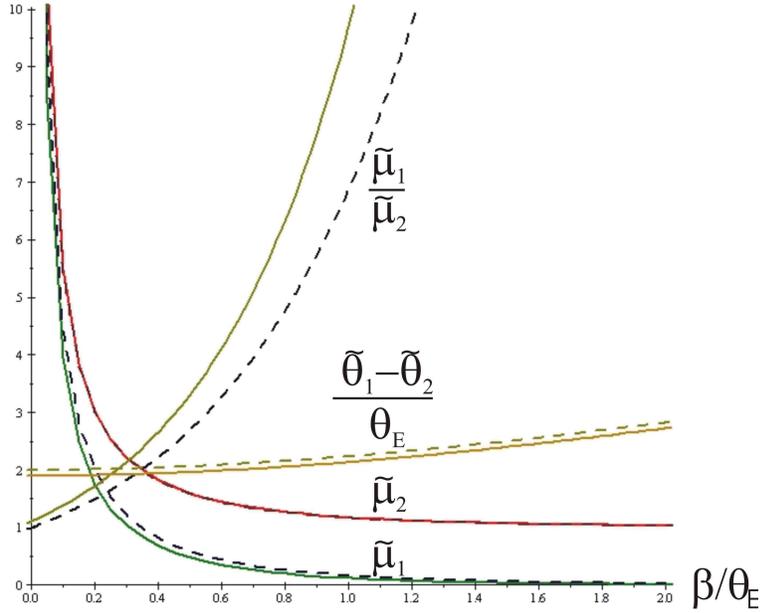}
\end{center}
\caption{ The image separations, the magnifications of the two images and
their ratio as functions of $\protect\beta /\protect\theta _{E}$ for the
perturbed case (solid lines) as compared to the Schwarzschild case (dashed
lines) for the parameter values $\protect\gamma \left( \bar{\protect\eta}-5%
\bar{\protect\varepsilon}^{2}\right) =10^{-1}\protect\theta _{E}^{3}$. From
top to bottom at $\protect\beta /\protect\theta _{E}=0.8$ they are the flux
ratio $\widetilde{\protect\mu }_{1}/\widetilde{\protect\mu }_{2}$, the image
separation $\widetilde{(\protect\theta }_{1}-\widetilde{\protect\theta }%
_{2})/\protect\theta _{E}$, the magnification of the primary image $%
\widetilde{\protect\mu }_{1}$ and finally the magnification of the secondary
image $\widetilde{\protect\mu }_{2}$. The largest effect can be seen on the
flux ratios.}
\label{mag1}
\end{figure}

We show in Fig \ref{mag1} the image separations, the magnifications of the
two images and their ratio for both the perturbed and Schwarzschild cases
for the case where $\gamma \left( \bar{\eta}-5\bar{\varepsilon}^{2}\right)
=10^{-1}\theta _{E}^{3}$. This figure is an analogue of Fig. 2.4. of Ref. 
\cite{SEF}.

The image separation slightly decreases\ in the perturbed case, with a
difference independent of $\beta $ between the tidal charged and the
Schwarzschild black holes. The primary image magnification is negligibly
affected, while the changes in the magnification of the secondary image are
more significant and lead to a lessening of its brightness. This effect can
be expected from the fact that the bending angle for the secondary image is
greater than that for the primary, and therefore creates a greater
sensitivity to changes in the the geometry. Finally, the differences in the
magnification ratios ($\tilde{\mu}_1/\tilde{\mu}_2$) for the two black holes
are most apparent as $\beta $ increases: the magnification ratio is
significantly larger in the perturbed case. From an observational point of
view such a measure should provide the best means to distinguish between the
two black hole geometries.

We have checked that in the situation where $\bar{\eta}-5\bar{\varepsilon}%
^{2}<0$, the changes with respect to the Schwarzschild black hole lensing
will be reversed. The image separation and the secondary image magnification
increase, while the magnification ratio decreases.

\section{Tidal charge dominated weak lensing to first order\label{1o_q}}

Under the assumptions B) of Section \ref{2o_lens} the lens equation (\ref%
{LE1}) reduces to%
\begin{equation}
0=\cos ^{2}\theta ~\left( \tan \left\vert \theta \right\vert -s\tan \beta
\right) +\bar{\eta}\frac{3\pi }{4}\frac{L}{D_{l}}\cot \left\vert \theta
\right\vert \left( \cot \left\vert \theta \right\vert +s\frac{L}{D_{l}}\tan
\beta \right) ~.
\end{equation}%
The expansion in $\theta $ and $\beta $ to leading order yields (assuming $%
\bar{\eta}=\mathcal{O}\left( \theta ^{3}\right) $, which turns out to be the
weak lensing condition for the tidal charge dominated black hole):%
\begin{equation}
0=\theta ^{3}-\beta \theta ^{2}+s\gamma \bar{\eta}~.  \label{LE_q}
\end{equation}%
$\allowbreak $Here again, without tidal charge there is no deflection ($%
\beta =\theta $).

The tidal charge causes a small deflection, which will be discussed in what
follows. In the process we need to identify real roots of the third rank
polynomial and give them in manifestly real form. We proceed as follows. The
discriminant of the third rank polynomial on the right hand side of Eq. (\ref%
{LE_q}) is 
\begin{equation}
\Delta =s\bar{\eta}\gamma \left( -\frac{\beta ^{3}}{27}+\frac{s\bar{\eta}%
\gamma }{4}\right) ~.
\end{equation}%
There is one real root when $\Delta >0$. This situation occurs when either $s%
\bar{\eta}<0$ or $s\bar{\eta}>4\beta ^{3}/27\gamma $. Otherwise, when $0<s%
\bar{\eta}\leq 4\beta ^{3}/27\gamma $ there are three real roots (two of
them being equal, when the equality holds). The case $s\bar{\eta}=0$ would
mean no deflection, thus it is excluded (we assume non-collinearity, $\beta
\neq 0$). We discuss the individual cases as follows:

\textit{Case} $s\bar{\eta}<0$. The polynomial (\ref{LE_q}) has one real root 
$\theta _{\eta 1}$, which is an image only if it obeys \textrm{sgn }$\theta
_{\eta 1}\equiv s=-$\textrm{sgn }$\bar{\eta}$. The real root (obtained by
the Cardano formula) 
\begin{equation}
\theta _{\eta 1}=\frac{\beta }{3}+\frac{2\beta }{3}\cosh \omega _{1}>\beta ~,
\label{hippoz}
\end{equation}%
is positive, thus $s=1$. Here we have introduced the notation%
\begin{equation}
\omega _{s}=\frac{1}{3}\textrm{arc}\cosh \left( s-\frac{27}{2}\frac{\gamma 
\bar{\eta}}{\beta ^{3}}\right) ~.
\end{equation}%
This image arises due to a negative tidal charge, which has the same type of
lensing effect as the mass would have in the Schwarzschild case (Fig \ref%
{fig1}.a).

We will identify the second image $\theta _{\eta 2}$, which forms below the
optical axis, by analyzing the remaining two cases. Both these cases obey $s%
\bar{\eta}>0$, thus \textrm{sgn }$\theta _{\eta 2}=s=$\textrm{sgn }$\bar{\eta%
}$. They include images formed below the optical axes ($s=-1$) only for
negative tidal charges (Fig \ref{fig1}.b); and above the optical axis ($s=1$%
) only for positive tidal charges (which generate a repulsive, scattering
effect, see Fig \ref{fig1}.c). As there are two sign changes in the
polynomial (\ref{LE_q}), from the generic theory of polynomials we expect an
even number of positive roots. We discuss these two cases in what follows.

\textit{Case} $s\bar{\eta}>4\beta ^{3}/27\gamma $. The polynomial (\ref{LE_q}%
) has one real root $\theta _{\eta 2}$. The Cardano formula gives a negative
root,%
\begin{equation}
\theta _{\eta 2}=\frac{\beta }{3}-\frac{2\beta }{3}\cosh \omega _{-1}\leq -%
\frac{\beta }{3}~,  \label{hipneg}
\end{equation}%
therefore $s=-1$ holds. This is the second image due to a negative tidal
charge.

\textit{Case} $0<s\bar{\eta}\leq 4\beta ^{3}/27\gamma $. The polynomial (\ref%
{LE_q}) has three real roots, one negative 
\begin{equation}
\theta _{\eta 2}^{^{\prime }}=\frac{\beta }{3}+\frac{2\beta }{3}\cos \varphi
_{-1}\in \frac{\beta }{3}[-1,0)~,  \label{har2}
\end{equation}%
($\cos \varphi _{-1}\in \lbrack -1,-1/2)$) and two positive%
\begin{eqnarray}
\theta _{\eta 2}^{^{\prime \prime }} &=&\frac{\beta }{3}+\frac{2\beta }{3}%
\cos \left( \varphi _{1}+\frac{2\pi }{3}\right) \in (0,\beta )~.
\label{har4} \\
\theta _{\eta 1}^{^{\prime \prime }} &=&\frac{\beta }{3}+\frac{2\beta }{3}%
\cos \left( \varphi _{1}-\frac{2\pi }{3}\right) \in (0,\beta )~.
\end{eqnarray}%
Here we have introduced the notation%
\begin{equation}
\varphi _{s}=\frac{1}{3}\arccos \left( 1-\frac{27}{2}s\frac{\gamma \bar{\eta}%
}{\beta ^{3}}\right) +\frac{2\pi }{3}.  \label{phis}
\end{equation}%
We remark, that $\theta _{\eta 2}^{^{\prime \prime }}<\theta _{\eta
1}^{^{\prime \prime }}$ on the whole range, with the exception of $\bar{\eta}%
=4\beta ^{3}/27\gamma $, where $\theta _{\eta 1}^{^{\prime \prime }}=\theta
_{\eta 2}^{^{\prime \prime }}=2\beta /3$.

The negative root $\theta _{\eta 2}^{^{\prime }}$ corresponds to a negative
tidal charge, which is the second image corresponding to $\theta _{\eta 1}$
in the range of parameters covered in this case. The positive roots $\theta
_{\eta 1}^{^{\prime \prime }},$ $\theta _{\eta 2}^{^{\prime \prime }}$
correspond to a positive tidal charge, which induces a scattering (a lensing
with $\delta <0$).

In the $\bar{\eta}\rightarrow 0$ limit, we expect the image at $\beta $.
Indeed, $\omega _{s}=0,~\varphi =2\pi /3$, thus $\theta _{\eta 1}\rightarrow
\beta $ and $\theta _{\eta 2}^{^{\prime }}\rightarrow 0$ for $\bar{\eta}<0$
(this being in perfect analogy with the behaviour of the Schwarzschild
images $\theta _{1,2}$ in the $\bar{\varepsilon}\rightarrow 0$ limit); while 
$\theta _{\eta 2}^{^{\prime \prime }}\rightarrow 0$ and $\theta _{\eta
1}^{^{\prime \prime }}\rightarrow \beta $ for $\bar{\eta}>0$.

\subsection{Summary of the image positions}

For negative tidal charge we obtained two images, $\theta _{\eta 1}$
appearing above the optical axis, and (depending on the magnitude of the
tidal charge) either $\theta _{\eta 2}$ or $\theta _{\eta 2}^{^{\prime }}$,
appearing below the optical axis. We remark here, that by the identity $%
\cosh x=\allowbreak \cos ix$ one can show that $\varphi =\pi -i\omega _{-1}$
holds, such that the two expressions can be shown to be identical, $\theta
_{\eta 2}^{^{\prime }}\equiv \theta _{\eta 2}$. However only one of the
expressions $\theta _{\eta 2}^{^{\prime }}$ and $\theta _{\eta 2}$ is
manifestly real, each in its domain of validity.

For positive tidal charge we have obtained two images $\theta _{\eta
1}^{^{\prime \prime }},$ $\theta _{\eta 2}^{^{\prime \prime }}$ of the type
represented on Fig \ref{fig1}(b). The possibility to have more then one such
trajectory for a given configuration of the source, lensing object and
observer is encoded in the fact that the deflection is stronger as we
approach the black hole. This is in perfect analogy with the scattering
produced by a negative mass Schwarzschild centre (a naked singularity),
discussed in Ref. \cite{Cramer}. Another similarity is the existence of an
upper limit $\bar{\eta}_{\max }=4\beta ^{3}/27\gamma $, which is the largest
value capable of producing scattered images.

\begin{figure}[tbp]
\begin{center}
\includegraphics[width=7cm]{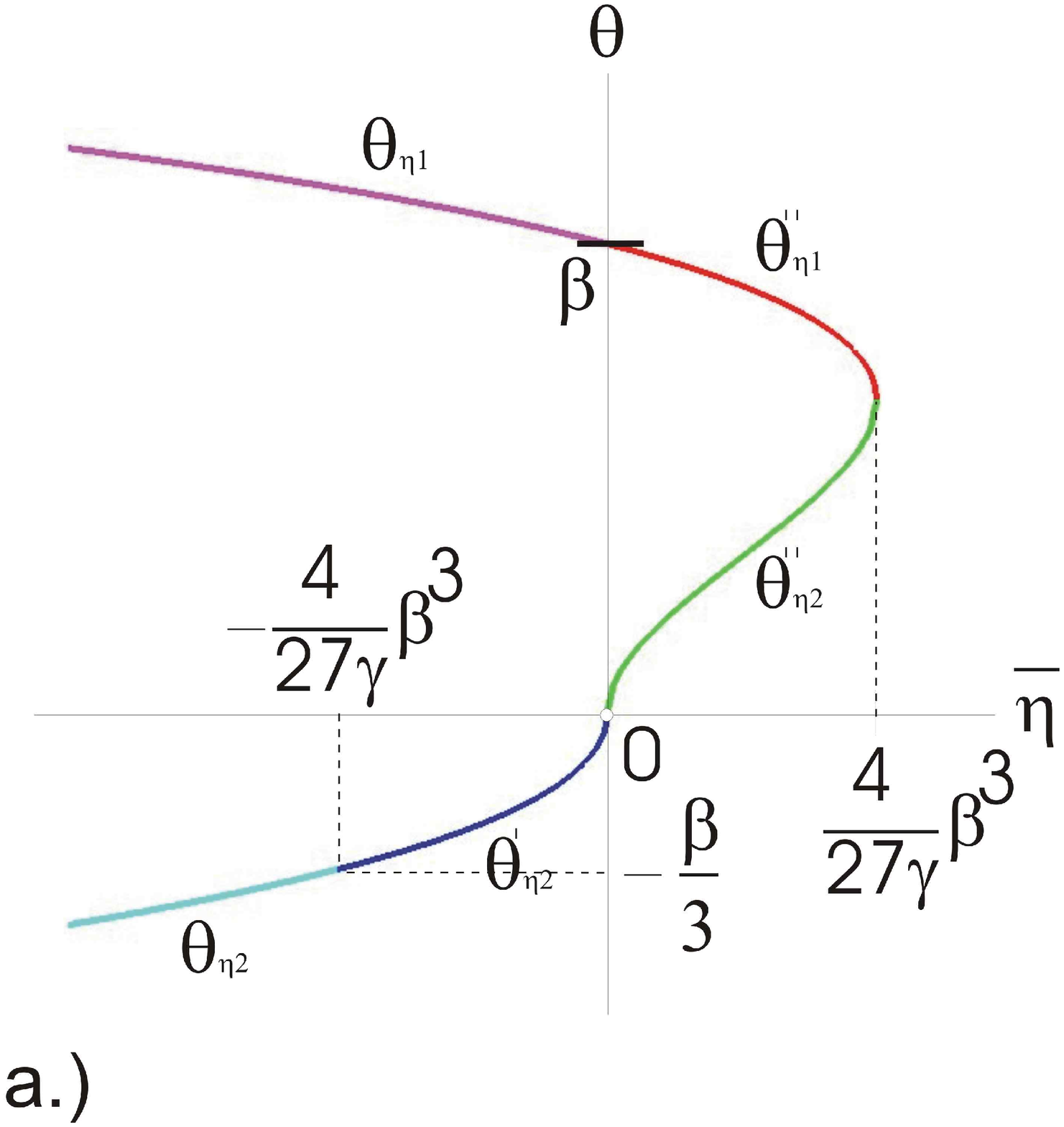} \vskip1cm %
\includegraphics[width=10cm]{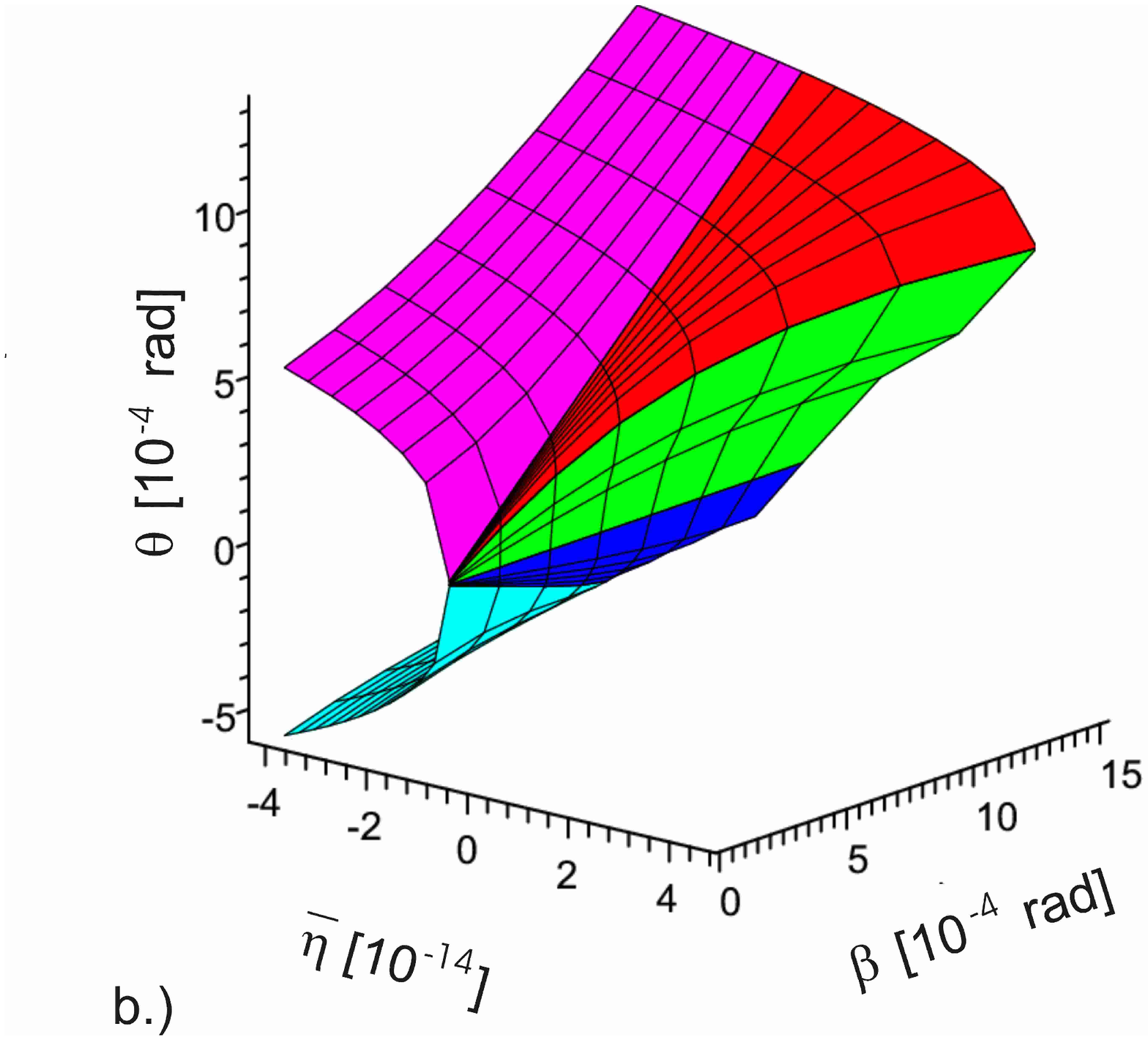}
\end{center}
\caption{ When the tidal charge dominates over the mass in the lensing,
there are still two images, but different from the Schwarzschild case. (a)
The position $\protect\theta $ (in units of $\protect\beta $) of the images
for tidal charge dominated black holes characterized by the parameter $\bar{%
\protect\eta}$. For negative tidal charge ($\bar{\protect\eta}<0$) there are
two images, situated above and below the optical axis. For positive tidal
charge ($\bar{\protect\eta}>0$) both images lie above the optical axis. The
images coincide for $\bar{\protect\eta}=4\protect\beta ^{3}/27\protect\gamma 
$. For each $\protect\beta $, the positive tidal charges with $\bar{\protect%
\eta}>4\protect\beta ^{3}/27\protect\gamma $ do not allow for any image. The
colours distinguish the images obtained as distinct analytic expressions,
which however generate a globally continuous curve. (b) $\protect\theta $ as
function of $\bar{\protect\eta}$ and $\protect\beta $. With decreasing $%
\protect\beta $, the images shrink accordingly. At $\protect\beta =0,~$the
angle $\protect\theta $ represents the angular radius of the Einstein ring
(therefore the $\protect\beta =0$ section of the surface is symmetric with
respect to $\protect\theta =0$).}
\label{fig3}
\end{figure}

The image locations for the different ranges of $\bar{\eta}$ are presented
in Fig \ref{fig3}.

\subsection{Einstein rings formed by tidal charged lenses without mass}

When $\beta =0$, Eq$.$ (\ref{LE_q}) becomes%
\begin{equation}
0=~\left\vert \theta \right\vert ^{3}+\bar{\eta}\gamma ~.
\end{equation}%
This has no solution for positive tidal charge. For negative tidal charge we
get the analogue of the Einstein ring at $\theta _{\eta E}=$ $\gamma
^{1/3}\left\vert \bar{\eta}\right\vert ^{1/3}.$

Due to the presence of $\beta $ in the denominators, the $\beta \rightarrow
0 $ limit cannot be obtained directly from the analytical expressions for $%
\theta _{\eta 1}$, $\theta _{\eta 2}$ and $\theta _{\eta 2}^{^{\prime }}$.
We have checked, using the l'Hospital rule that $\lim_{\beta \rightarrow
0}\theta _{\eta 1}=$ $-\lim_{\beta \rightarrow 0}\theta _{\eta
2}=-\lim_{\beta \rightarrow 0}\theta _{\eta 2}^{^{\prime }}=$ $\theta _{\eta
E}$.

\subsection{Magnification factors}

The magnification factor for each of the images discussed above, namely; $%
\theta _{\eta 1,2}$, $\theta _{\eta 2}^{\prime }\,$\ and $\theta _{\eta
1,2}^{^{\prime \prime }}$, respectively are found to be%
\begin{eqnarray}
\mu _{\eta 1,2} &=&\frac{2}{9}\sqrt{\frac{27 \theta _{\eta E}^3}{27 \theta
_{\eta E}^3\mp 4\beta ^{3}}}\left( 2\cosh \omega _{\pm 1}\pm 1\right) \sinh
\omega _{\pm 1}+\frac{\left( 1\pm 2\cosh \omega _{\pm 1}\right) ^{2}}{9}~, \\
\mu _{\eta 2}^{\prime } &=&\frac{-2}{9}\sqrt{\frac{-27 \theta _{\eta E}^3}{%
27 \theta _{\eta E}^3 +4\beta ^{3}}}\left( 2\cos \varphi _{-1}+1\right) \sin
\varphi _{-1}+\frac{\left( 1+2\cos \varphi _{-1}\right) ^{2}}{9}~, \\
\mu _{\eta 1,2}^{^{\prime \prime }} &=&\pm \frac{2}{9}\sqrt{\frac{27 \theta
_{\eta E}^3}{-27 \theta _{\eta E}^3+4\beta ^{3}}}\left[ 2\cos \left( \varphi
_{1}\mp \frac{2\pi }{3}\right) +1\right] \sin \left( \varphi _{1}\mp \frac{%
2\pi }{3}\right)  \nonumber \\
&&\pm \frac{1}{9}\left[ 1+2\cos \left( \varphi _{1}\mp \frac{2\pi }{3}%
\right) \right] ^{2}~,
\end{eqnarray}%
$\allowbreak $

\begin{figure}[tbp]
\begin{center}
\includegraphics[width=10cm]{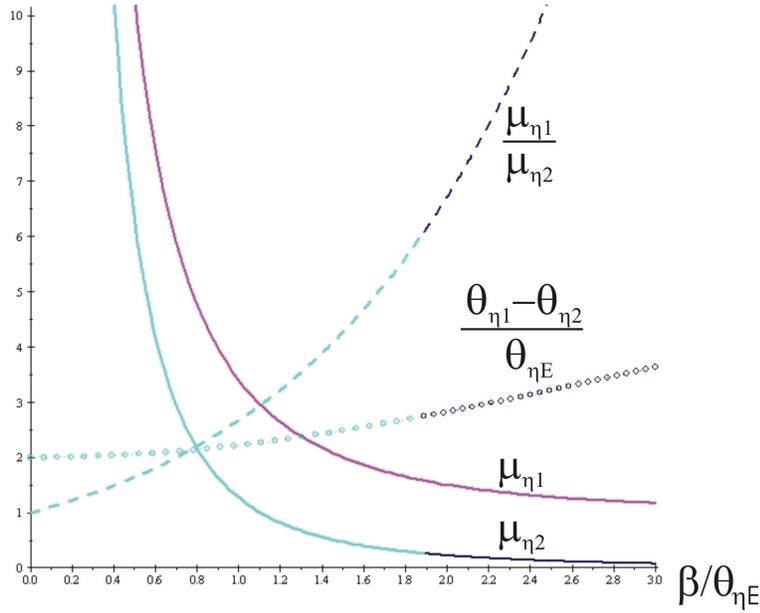}
\end{center}
\caption{ Image separations and magnifications for a negative tidal charge
dominated black hole, as functions of $\protect\beta /\protect\theta _{%
\protect\eta E}$. The upper and lower solid curves plot the primary and
secondary image magnification factors, respectively; their ratio is the
dashed curve; and the dotted curve is the image separation. A color change
at $3/4^{1/3}$ represents the change in functional form from $\protect\theta %
_{\protect\eta 2}$ (left) to $\protect\theta _{\protect\eta 2}^{\prime }$
(right)$.$The colours match those of Fig \protect\ref{fig3}.}
\label{mag2}
\end{figure}

In Fig \ref{mag2} we have represented the normalized image separations in
units of $\theta _{\eta E}$, the magnification factors and the flux ratios
for a negative tidal charge as function of $\beta /\theta _{\eta E} $. The
image $\theta _{\eta 2}$ arises for $\bar{\eta}\leq -4\beta ^{3}/27\gamma $
(thus $-\theta _{\eta E}\leq -4^{1/3}\beta /3$), such that $\beta /\theta
_{\eta E}\leq 3/4^{1/3}=\allowbreak 1.8899$ while the image $\theta _{\eta
2}^{\prime }$ for $-4\beta ^{3}/27\gamma \leq \bar{\eta}\leq 0$ (thus $%
-4^{1/3}\beta /{3}\leq -\theta _{\eta E}\leq 0$), such that $\beta /\theta
_{\eta E}\geq 3/4^{1/3}$. By contrast, the image $\theta _{\eta 1}$ arises
for any $\bar{\eta}\leq 0$. Therefore we have plotted $\left( \theta _{\eta
1}-\theta _{\eta 2}\right) /\theta _{\eta E}$, $\mu _{\eta 1}$, $\mu _{\eta
2}$, and $\mu _{\eta 1}/\mu _{\eta 2}$ in the range $\beta /\theta _{\eta
E}\in \left[ 0,3/4^{1/3}\right] $ while $\left( \theta _{\eta 1}-\theta
_{\eta 2}^{\prime }\right) /\theta _{\eta E}$, $\mu _{\eta 1}$, $\mu _{\eta
2}^{\prime }$, and $\mu _{\eta 1}/\mu _{\eta 2}^{\prime }$ in the range $%
\beta /\theta _{\eta E}\geq 3/4^{1/3}$. 

As for Schwarzschild case, the image separation and the flux ratio are
monotonically increasing functions of the source angle $\beta $, while the
magnification factors decrease with increasing $\beta $, and diverge for $%
\beta \rightarrow 0$. Also the flux ratio goes to unity as $\beta
\rightarrow 0$. The most obvious difference between the tidal charge
dominated spacetime and the Schwarzschild case is that for a fixed value of
the image separation, the magnification factors are significantly increased
in the former. Unfortunately a measurement of the individual magnification
factors requires a knowledge of the unlensed source brightness. However if
the images can be resolved to obtain their angular separation and individual
brightnesses one can compare the ratio of the magnification factors as a
function of the image separation normalized to the Einstein angle for the
system. This has the advantage of not having to refer to the unlensed source
brightness and normalizes the image separation using the characteristic
lensing parameters.

\begin{figure}[tbp]
\begin{center}
\includegraphics[width=10cm]{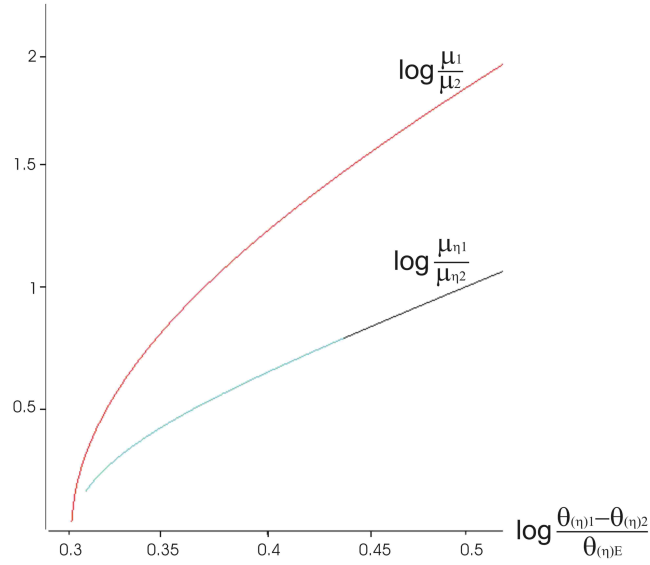}
\end{center}
\caption{ The ratio of the magnification factor of the primary and secondary
images as function of the image separation normalized to the Einstein angle,
on log-log scale, for the tidal charge dominated black hole and
Schwarzschild black hole. The tidal charged case is distinguished by the use
of the $\protect\eta$-subscript.}
\label{magsep}
\end{figure}

Figure \ref{magsep} plots the logarithm of the ratio of the primary
magnification to the secondary magnification as a function of the logarithm
of the image separation divided by the Einstein angle. For image separations
greater than about 2.5 times the Einstein angle, the ratio of the
magnification factors for each image obeys a power law relationship. Since
this leads to an independence of scaling, we suggest that observations of
image brightnesses and image separations should be able to distinguish
easily between the standard Schwarzschild spacetime and that governed by
higher-dimensional Weyl curvature effects that induce the tidal charge.
Given that the magnification of the secondary image produced by the
Schwarzschild lens is significantly reduced, the ratio $\mu _{1}/\mu _{2}$
is much larger than that for the tidally charged lens. Thus for large image
separations one has the relation: 
\[
\mu _{1}/\mu _{2}\approx \lbrack \Delta \theta /\theta _{E}]^{\kappa } 
\]%
In the case of Schwarzschild lensing $\kappa =6.22\pm .15$ whereas for the
tidally charged black hole lensing one obtains $\kappa _{\eta }=2.85\pm .25$%
, which gives two completely different power law behaviours.

Therefore given a large enough number of measurements of image separations
and image brightnesses (as well as a knowledge of the characteristics of the
lensing object) such a relationship should provide a very good observational
signature that might distinguish between the lensing behaviour by the two
types of black holes. 

\begin{figure}[tbp]
\begin{center}
\includegraphics[width=9cm]{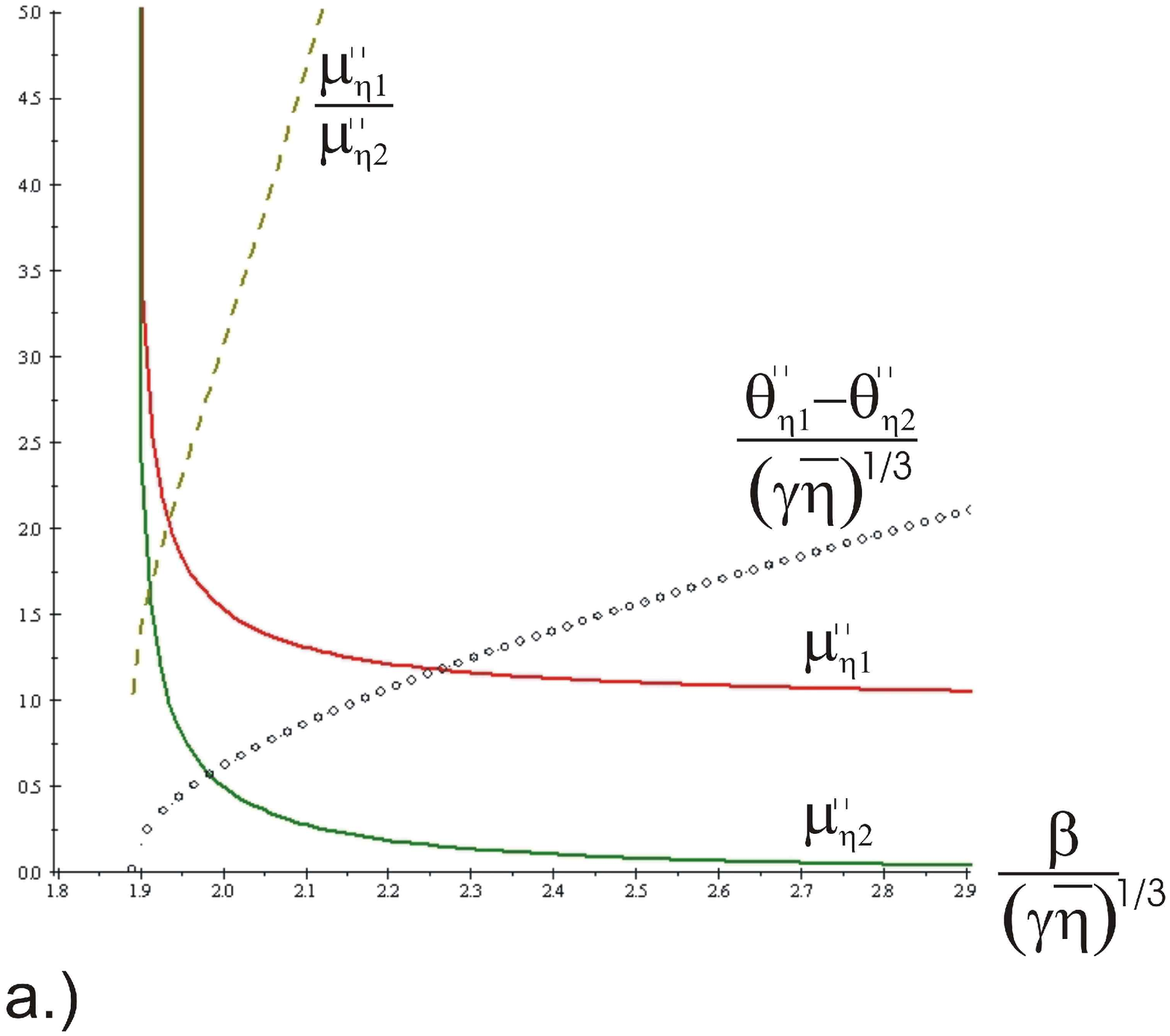} \vskip1cm %
\includegraphics[width=9cm]{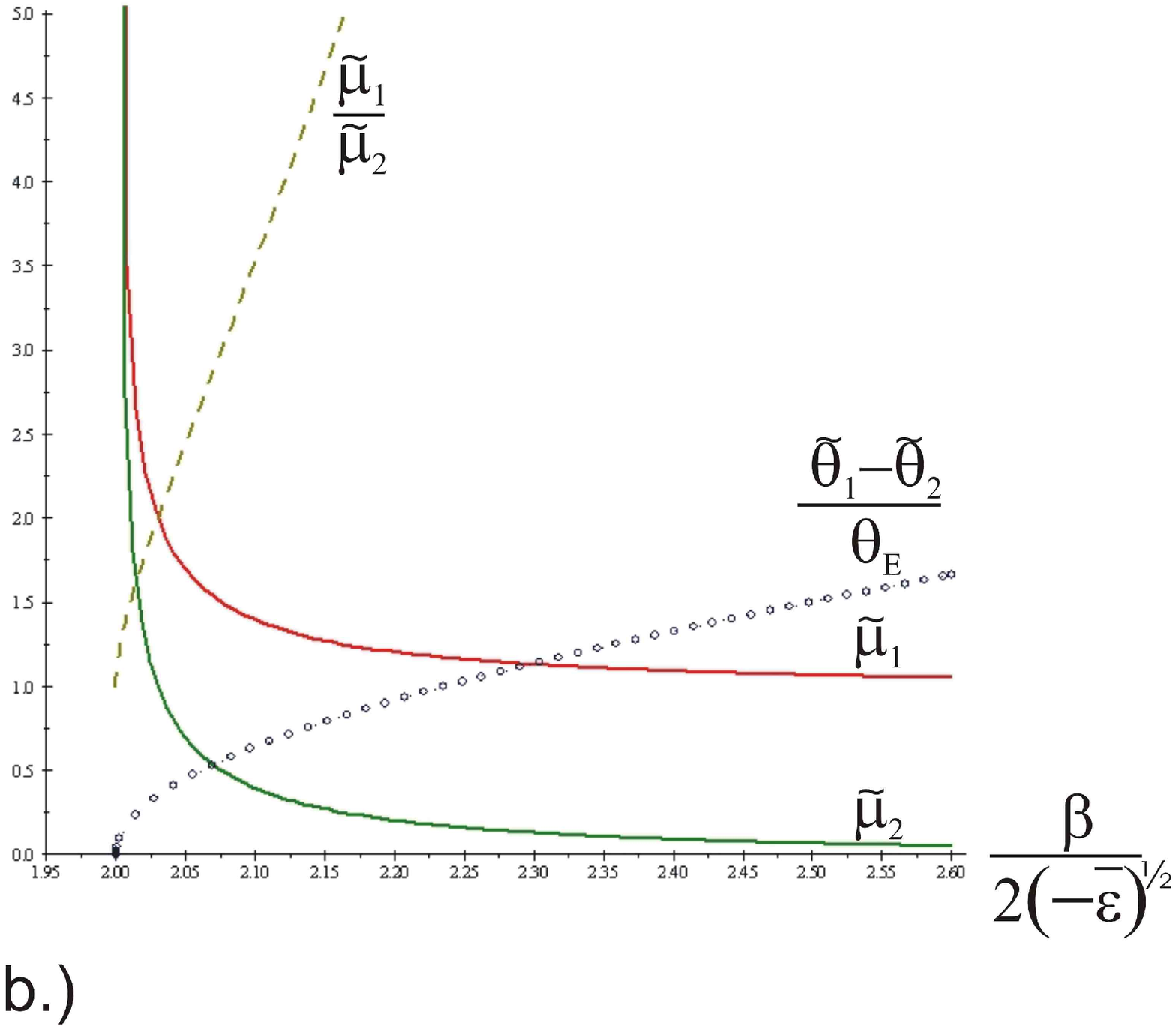}
\end{center}
\caption{ Image separations and magnifications for a positive tidal charge
dominated black hole (a), as functions of $\protect\beta /\left( \protect%
\gamma \bar{\protect\eta}\right) ^{1/3}$. The upper and lower solid curves
plot the primary and secondary image magnification factors, respectively;
their ratio is the dashed curve; and the dotted curve is the image
separation. The colours match those of Fig \protect\ref{fig3}. The negative
mass Schwarzschild geometry produces similar lensing effects (b), in terms
of the independent variable $\protect\beta /2\left( -\bar{\protect\varepsilon%
}\right) ^{1/2}$.}
\label{mag3}
\end{figure}

In Fig \ref{mag3} we have represented the image separations $\left( \theta
_{\eta 1}^{^{\prime \prime }}-\theta _{\eta 2}^{^{\prime \prime }}\right)
/\left(\gamma \bar{\eta}\right)^{1/3} $, magnification factors $\mu _{\eta
1,2}^{^{\prime \prime }}$ and flux ratios $\mu _{\eta 1}^{^{\prime \prime
}}/\mu _{\eta 2}^{^{\prime \prime }}$ for a positive tidal charge as
function of $\beta /\left(\gamma \bar{\eta}\right)^{1/3} $. As the allowed
positive tidal charged parameter range is $0\leq \bar{\eta}\leq 4\beta
^{3}/27\gamma $ (thus $0\leq \left(\gamma \bar{\eta}\right)^{1/3} \leq
4^{1/3}\beta /{3}$), the range of the variable $x=\beta / \left(\gamma \bar{%
\eta}\right)^{1/3} $ is restricted to $x\geq 3/4^{1/3}$. This is very
similar to the negative mass Schwarzschild lensing, also shown on Fig \ref%
{mag3}.

\section{Tidal charge dominated weak lensing to second order\label{2o_q}}

In this section we follow the method presented in section \ref{2o_m} for
obtaining the second order correction to the tidal charge dominated light
deflection. Thus we go to second order in $\bar{\eta}$. The terms $\bar{%
\varepsilon}^{2}$ and$\ \bar{\varepsilon}\bar{\eta}$ can be dropped from the
lens equation (\ref{LE1}), being considered of higher order. Thus the lens
equation (\ref{LE1}) reduces to%
\begin{eqnarray}
0 &=&\cos ^{2}\theta \left( \tan \theta -\tan \beta \right) -4\bar{%
\varepsilon}\cos \theta \left( \frac{4\gamma }{3\pi }\tan \beta +\cot \theta
\right)  \nonumber \\
&&+\gamma \bar{\eta}\cot \theta \left( s\cot \theta +\frac{4\gamma }{3\pi }%
\tan \beta \right)  \nonumber \\
&&-\frac{\gamma ^{3}}{9\pi ^{2}}\bar{\eta}^{2}\frac{\cos \theta }{\sin
^{3}\theta }\left[ \frac{4\gamma }{3\pi }\tan \beta \left( 6\pi \cot \theta
-35s\right) -6\pi -35s\cot \theta \right] ~.
\end{eqnarray}%
By taking $\bar{\eta}=\mathcal{O}\left( \theta ^{3}\right) $ in accordance
with the reasoning of section \ref{1o_q}, and a higher order mass parameter $%
\bar{\varepsilon}=\mathcal{O}\left( \theta ^{\geq 4}\right) $, the expansion
in $\theta $ and $\beta $ yields

\begin{eqnarray}
0 &=&\theta -\beta +{\frac{s\gamma \bar{\eta}}{{\theta }^{2}}+}\mathcal{U}%
\left( \bar{\varepsilon},\bar{\eta},s,\theta \right) ~,  \nonumber \\
\mathcal{U} &=&-\frac{35s\gamma ^{3}}{9\pi ^{2}}\frac{\bar{\eta}^{2}}{\theta
^{4}}+\mathcal{U}_{3}{~,}  \label{U23}
\end{eqnarray}%
where $\mathcal{U}_{3}$ represents third order terms in $\theta $ as follows 
\begin{eqnarray}
\mathcal{U}_{3} &=&-\frac{2\theta +\beta }{3}\left( \theta -\beta \right)
^{2}-{\frac{4\bar{\varepsilon}}{\,\theta }}  \nonumber \\
&&-\,{\frac{2s\gamma \bar{\eta}\,}{3}}+{\frac{4{\gamma }^{2}\beta \,\bar{\eta%
}\,}{3s\pi \,\theta }}-{\frac{{2\gamma }^{3}\bar{\eta}^{2}}{3\pi \,{\theta }%
^{3}}\ }+{\frac{8\,{\gamma }^{4}\beta \bar{\eta}^{2}}{9{\pi }^{2}{\theta }%
^{4}}~.}  \label{U3}
\end{eqnarray}%
According to our assumptions the leading order term due to the mass enters
here, therefore we need to keep all other terms of this order. Note that
some of these terms differ from the respective terms arising from the
Virbhadra-Ellis lens equation (\ref{lens_Ellis}), as can be checked from the
Appendix.

We also remark, that without mass and tidal charge there is no deflection at
all ($\theta =\beta $ is a solution when $\bar{\varepsilon}=\bar{\eta}=0$),
as expected.

The term $\mathcal{U}$ represents a perturbation of the tidal charge
dominated weak lensing discussed in the previous section, Eq. (\ref{LE_q})
and is of $\mathcal{O}\left( \theta ^{\geq 2}\right) $. We look for
solutions therefore in the form 
\begin{equation}
\widetilde{\theta }=\theta \left[ 1+\mathcal{T}_{\eta }\left( \bar{%
\varepsilon},\bar{\eta},s,\beta \right) \right] ~.
\end{equation}%
In what follows we will show that the correction term $\mathcal{T}_{\eta }$
is $\mathcal{O}\left( \theta \right) $. 

\subsection{Positive $\bar{\protect\eta}$}

We look for the corrections $\left( \mathcal{T}_{\eta }\right) _{\pm }\left(
\theta _{\eta }\right) _{\pm }$ of the images $\left( \theta _{\eta }\right)
_{\pm }$ located at (\ref{hippoz}), (\ref{hipneg}). The solutions are%
\begin{equation}
\left( \mathcal{T}_{\eta }\right) _{\pm }=\frac{\left( \theta _{\eta
}\right) _{\pm }^{2}\mathcal{U}\left( \bar{\varepsilon},\bar{\eta},s,\left(
\theta _{\eta }\right) _{\pm }\right) }{{2\gamma \bar{\eta}}-s\left( \theta
_{\eta }\right) _{\pm }^{3}}~.
\end{equation}%
{\ }There are two images, located at $\left( \widetilde{\theta }_{\eta
}\right) _{\pm }=\left( \theta _{\eta }\right) _{\pm }\left[ 1+\left( 
\mathcal{T}_{\eta }\right) _{\pm }\right] .$

\subsection{Negative $\bar{\protect\eta}$}

We look for the corrections $\left( \mathcal{T}_{\eta }\right) _{1.2}\left(
\theta _{\eta }\right) _{1,2}$ of the images $\left( \theta _{\eta }\right)
_{1,2}$ located at (\ref{hippoz}), (\ref{hipneg}). The solutions are%
\begin{equation}
\left( \mathcal{T}_{\eta }\right) _{1.2}=\frac{\left( \theta _{\eta }\right)
_{1,2}^{2}\mathcal{U}\left( \bar{\varepsilon},\bar{\eta},s,\left( \theta
_{\eta }\right) _{1,2}\right) }{{2\gamma \bar{\eta}}-s\left( \theta _{\eta
}\right) _{1,2}^{3}}~.
\end{equation}%
{\ }There are two images, located at $\left( \widetilde{\theta }_{\eta
}\right) _{1,2}=\left( \theta _{\eta }\right) _{1,2}\left[ 1+\left( \mathcal{%
T}_{\eta }\right) _{1.2}\right] .$

\section{Concluding remarks}

Using simple geometric relations we derived a generic lens equation, Eq. (%
\ref{LE}) for weak lensing. This formula is more accurate than the
Virbhadra-Ellis lens equation (as it contains no approximations of
trigonometric expressions), but reduces to it in a proper limit (differences
are to be expected in asymmetric source and observer distances with respect
to the lens).

We have applied our lens equation in the discussion of the weak lensing by
tidal charged black holes, to both first and second order accuracy in the
black hole parameters, when either the mass or the tidal charge dominates.
We have carried on expansions in the small mass and tidal charge parameters,
as well as in the angles spanned by the real and apparent positions of the
sources with the optical axis. In the Appendix we have investigated the
differences between the predictions of the two lens equations. The
predictions of our lens equation (\ref{LE}) and of the Virbhadra-Ellis lens
equation (\ref{lens_Ellis}) coincide in most of the cases we consider, with
the notable exception of the tidal charge dominated black hole lensing
discussed to a higher accuracy in Section \ref{2o_q}, where the
Virbhadra-Ellis lens equation would not predict (or would predict with
different coefficients) some of the higher order terms.

Although Solar System tests lead to the expected result, that light
deflection by the Sun is due to its mass, and leaves room but for a
minuscule tidal charge \cite{tidalDeflection}, \cite{BoehmerHarkoLobo}, it
cannot be excluded that tidal charge dominated black holes could exist on
the brane, as the tidal charge is an imprint of the Weyl curvature of the
higher-dimensional space-time, which remains unspecified for the tidal
charged black hole.

In the case of mass dominated weak lensing, we found that the position of
the images is similar to the Reissner-Nordstr\"{o}m black hole lensing,
discussed in Ref. \cite{Sereno}. We have analyzed, how the image
separations, the magnification factors and the flux ratios are modified as
compared to the Schwarzschild lensing by the perturbations arising from
second order mass and linear tidal charge contributions ($\bar{\varepsilon}%
^{2}$ and $\bar{\eta}$, respectively). The most striking modification
appears in the ratio of the magnification factors (the flux ratio), shown on
Fig \ref{mag1}, which can be either increased or decreased, depending on the
sign of $\bar{\eta}-5\bar{\varepsilon}^{2}$.

When the tidal charge dominates in the lensing behaviour, the situation is
different. The case of positive tidal charge resembles the lensing
properties of a negative mass Schwarzschild spacetime \cite{Cramer}.

Black holes with negative tidal charge are however favoured by strengthening
and confining gravity to the brane and also by thermodynamic considerations 
\cite{tidalThermo}. In the case of a dominant negative tidal charge the
lensing properties are similar to those of a positive mass Schwarzschild
black hole, where the similarity is only in the number of images lying above
or below the optical axis. The actual location of the images is different
and this fact is summarized in Fig \ref{fig3}, which is one of the main
results of this paper.

Finally the power law dependence of the ratio of the magnification factors
on the separation of the images provides an means for observing the
differences between the Schwarzschild and tidal charged black holes. Given
that the next generation of radio telescopes will easily be able to resolve
images to less than milli-arcsecond accuracy, the different rates at which
the ratio of the brightness changes should be able to provide a significant
observational signature to constrain the Weyl curvature as a substitute for
dark matter.

\section*{Acknowledgements}

ZsH and L\'{A}G were supported by the Hungarian Scientific Research Fund
(OTKA) grants nos. 69036 and 81364, respectively.

\appendix

\section{Comparing our lens equation with the Virbhadra-Ellis lens equation}

In this Appendix we present the explicit form of our lens equation (\ref{LE}%
) and of the Virbhadra-Ellis lens equation (\ref{lens_Ellis}), both to
fourth order in $\theta $, adopting the minimal assumptions $\bar{\varepsilon%
}=\mathcal{O}\left( \theta ^{\geq 2}\right) $ and $\bar{\eta}=\mathcal{O}%
\left( \theta ^{\geq 3}\right) $, which cover all cases considered in the
main text. These particular cases can be recovered by shifting the $\theta $%
-order of the parameters $\bar{\varepsilon}$ and $\bar{\eta}$, as described
at the beginning of each section, and dropping the terms, which fall beyond
the desired accuracy.

The lens equation (\ref{LE}) in detail reads:%
\begin{equation}
0=L_{0}^{HGH}+\bar{\varepsilon}L_{10}^{HGH}+\bar{\eta}L_{01}^{HGH}+\bar{%
\varepsilon}^{2}L_{20}^{HGE}+\bar{\varepsilon}\bar{\eta}L_{11}^{HGH}+\bar{%
\eta}^{2}L_{02}^{HGH}~,  \label{eqLE}
\end{equation}%
with the coefficients%
\begin{eqnarray}
L_{0}^{HGH} &=&\left( \theta -\beta \right) \left[ 1-\frac{\left( 2\theta
+\beta \right) \left( \theta -\beta \right) }{3}\right] {\ },  \nonumber \\
L_{10}^{HGH} &=&\,\,~-{\frac{4\,}{\theta }}+\frac{2}{3}\left( 5\,\theta \,{\ 
}-{\frac{8\gamma \beta \,}{\pi }}\right) ~,  \nonumber \\
L_{01}^{HGH} &=&{\frac{s\,\gamma }{{\theta }^{2}}}-\frac{2s{\gamma }}{3}%
\left( 1{\,}-{\frac{2{\gamma }\beta \,}{\pi \,\theta }}\right) ~,  \nonumber
\\
L_{20}^{HGE} &=&-{\frac{5s\gamma }{{\theta }^{2}}}-\frac{32\gamma }{3\pi
\,\theta }\left( {1}~-\,{\frac{4{\gamma }\beta }{3{\pi \theta }}}\right) +%
\frac{10s\gamma }{3}\left( {1}-{\frac{2{\gamma }\beta \,}{\pi \,\theta }}%
\right) ~,  \nonumber \\
L_{11}^{HGH} &=&{\frac{256{\gamma }^{2}}{9{\pi }^{2}{\theta }^{3}}}+\frac{16s%
{\gamma }^{2}}{3\pi \,{\theta }^{2}}\left( {1}-{\frac{4{\gamma }\beta \,}{3{%
\pi \theta }}}\right) -{\frac{128{\gamma }^{2}}{9{\pi }^{2}\theta }}\left( 1-%
{\frac{8{\gamma }\beta \,}{3{\pi \theta }}}\right) ~,  \nonumber \\
L_{02}^{HGH} &=&-{\frac{35s{\gamma }^{3}}{9{\pi }^{2}{\theta }^{4}}}-\frac{2{%
\gamma }^{3}}{3\pi \,{\theta }^{3}}\left( {1}-{\frac{4{\gamma }\beta \,}{3{%
\pi \theta }}}\right) +\frac{35{s\gamma }^{3}}{27{\pi }^{2}{\theta }^{2}}%
\left( {1}-{\frac{4{\gamma }\beta \,}{{\pi \theta }}}\right) ~.
\end{eqnarray}%
The Virbhadra-Ellis lens equation (\ref{lens_Ellis}) gives:%
\begin{equation}
0=L_{0}^{VE}+\bar{\varepsilon}L_{10}^{VE}+\bar{\eta}L_{01}^{VE}+\bar{%
\varepsilon}^{2}L_{20}^{VE}+\bar{\varepsilon}\bar{\eta}L_{11}^{VE}+\bar{\eta}%
^{2}L_{02}^{VE}~.  \label{eqVE}
\end{equation}%
The coefficients in (\ref{eqVE}) are:%
\begin{eqnarray}
L_{0}^{VE} &=&\left( \theta \,-\beta \right) \left( 1+\frac{\theta
^{2}+\beta ^{2}+\theta \beta }{3}\right) ~,  \nonumber \\
L_{10}^{VE} &=&-\,{\frac{4\,}{\theta }}-\frac{14\theta }{3}~,  \nonumber \\
L_{01}^{VE} &=&~{\frac{s\,\gamma }{{\theta }^{2}}}+\,{\frac{4s\gamma \,}{3}~}%
,  \nonumber \\
L_{20}^{VE} &=&-{\frac{5s\gamma }{{\theta }^{2}}}+{\frac{64\gamma }{3\pi
\theta }}-\,{\frac{20s\gamma }{3}}~,  \nonumber \\
L_{11}^{VE} &=&{\frac{256{\gamma }^{2}}{9{\pi }^{2}{\theta }^{3}}}-{\frac{32s%
{\gamma }^{2}}{3\pi \,{\theta }^{2}}}+{\frac{128{\gamma }^{2}}{3{\pi }%
^{2}\theta }}~,  \nonumber \\
L_{02}^{VE} &=&~-{\frac{35s{\gamma }^{3}}{9{\pi }^{2}{\theta }^{4}}+\frac{4{%
\gamma }^{3}}{3\pi \,{\theta }^{3}}}-{\frac{175s{\gamma }^{3}}{27{\pi }^{2}{%
\theta }^{2}}~}.
\end{eqnarray}%
One can see that all coefficients are different in the two approaches by
terms, which in the lens equations are of order $\mathcal{O}\left( \theta
^{\geq 3}\right) $. We conclude, that the two lens equations agree only at $%
\theta ^{2}$ order.

\end{document}